\documentclass[final,3p,times,twocolumn]{elsarticle}

\usepackage{amssymb}
\usepackage{amsmath}
\usepackage{multirow}
\usepackage{array}
\usepackage{float}
\usepackage{xhfill}
\usepackage{tabularx, booktabs}
\usepackage{adjustbox}

\usepackage{hyperref}
\usepackage{cleveref}

\journal{Information Fusion}

\begin{document}

\begin{frontmatter}



\title{Hybrid Quantum-Classical Convolutional Neural Networks for Image Classification in Multiple Color Spaces}

\author[1]{Kwok-Ho Ng}
\author[1]{Tingting Song}
\author[1]{Zhiquan Liu}

\affiliation[1]{organization={College of Cyber Security, Jinan University},
            addressline={601 Huangpu Avenue West}, 
            city={Guangzhou},
            postcode={510632}, 
            state={Guangdong},
            country={China}}

\begin{abstract}
The growing complexity and scale of image processing tasks challenge classical convolutional neural networks (CNNs) with high computational costs. Hybrid quantum-classical convolutional neural networks (HQCNNs) show potential to improve performance by accelerating processing speed, enhancing classification accuracy, and reducing model parameters, though studies have primarily focused on the RGB color space. 
However, the effectiveness of HQCNNs in non-RGB color spaces, such as Lab, YCrCb, and HSV, remains largely unexplored.
We propose an HQCNN to evaluate image classification across diverse color spaces.
The HQCNN integrates parameterized quantum circuits (PQCs) with a classical CNN, leveraging quantum entanglement and trainable gates to enhance expressiveness across varied color representations.
We assess performance on MNIST, CIFAR-10, EuroSAT, and SAT-4 datasets.
Experimental results demonstrate that the HQCNN outperforms the classical CNN across all tested color spaces for the ten-class MNIST task, achieving a best accuracy of $94.3\%$ in Lab compared to $92.8\%$ in RGB for the CNN, with superior performance on other datasets in various color spaces.
These findings highlight the potential of non-RGB color spaces and optimized PQC designs to improve classification performance. 
We provide new insights for advancing hybrid quantum-classical computer vision through optimized PQC architectures and diverse color space applications.

\end{abstract}

%



\end{frontmatter}



\section{Introduction}
\label{sec:intro}

Computer image classification is one of the core research areas in computer vision, with widespread applications across multiple fields. Notable examples include object classification in autonomous driving~\cite{1_de1997road,2_turay2022toward}, disease classification using computed tomography images in medicine~\cite{3_song2017using,4_pham2020comprehensive}, defect detection in industrial settings~\cite{5_ren2022state,6_singh2023automated}, and product identification and classification in agriculture~\cite{7_yang2022development,8_paymode2022transfer}. Advances in classification accuracy and algorithmic development continue to drive rapid and efficient progress across these disciplines.
Convolutional neural networks (CNNs) have been a cornerstone of computer vision tasks, particularly excelling in image recognition over the past decade~\cite{9_rawat2017deep,10_voulodimos2018deep}. Their hierarchical structure is designed to extract image features efficiently. 
The evolution of model scale and the emergence of innovative architectures, such as ResNet~\cite{11_he2016deep} and DenseNet~\cite{12_huang2017densely}, suggest that CNNs remain at the forefront of image classification methodologies.

However, classical algorithms face significant computational challenges when processing high-dimensional data, a bottleneck that becomes especially apparent in more complex tasks.
Quantum information and computing technologies are advancing rapidly, with quantum algorithms promising exponential performance improvements~\cite{13_jozsa2003role,14_childs2003exponential,15_babbush2021focus}.
These advancements have the potential to substantially enhance performance in computer vision and other applications requiring high-dimensional data processing.
Additionally, quantum neural networks (QNNs), a subset of quantum algorithms, are being actively developed to integrate quantum computing with machine learning, potential offering advantages~\cite{16_schuld2014quest}. 
Research by Gil-Fuster et al.~\cite {17_gil2024understanding} indicates that quantum machine learning may achieve effective generalization with minimal training data.
Recent studies~\cite{17_1gonon2025universal} further demonstrate that classical functions can be approximated using parameterized quantum circuits (PQCs), consistent with the universal approximation theorem.

Quantum convolutional neural networks (QCNNs) are quantum algorithms inspired by classical CNNs. 
Proposed by Cong et al.~\cite{18_cong2019quantum}, their structure mirrors that of CNNs, incorporating convolutional, pooling, and fully connected layers to extract features and process information from quantum states.
By combining the layered architecture of classical convolutional networks with the high-dimensional capabilities of quantum computing, QCNNs may offer notable advantages in handling complex data.
Notably, QCNNs can utilize a small number of variational parameters for efficient training, suggesting their potential for feature extraction and classification in image processing tasks~\cite{17_gil2024understanding}. 
Moreover, QCNNs have been shown to avoid barren plateaus~\cite{20_pesah2021absence}, distinguishing them from other QNNs by remaining trainable even under random initialization, thus avoiding the vanishing gradient issues prevalent in other quantum models.

Currently, QCNN applications in image classification primarily focus on single-channel datasets, such as MNIST, where they achieve high classification accuracy with significantly fewer parameters than traditional models~\cite{31_senokosov2024quantum}. 
However, their application to color image datasets remains limited, with most research focusing on the RGB color space. The exploration of alternative color spaces, such as Lab, YCrCb, and HSV, has been relatively underexplored.

In contrast, classical CNN research has extensively investigated various color spaces. 
For example, the Lab color space excels in identifying plant pests~\cite{21_khanramaki2021citrus}, while YCrCb outperforms RGB in image forgery detection~\cite{22_phan2019preserving}, and the HSV color model improves tomato ripeness identification in complex environments for automated harvesting~\cite{23_moreira2022benchmark}. 

\begin{figure}[htbp]
  \centering
   \includegraphics[width=0.9\linewidth]{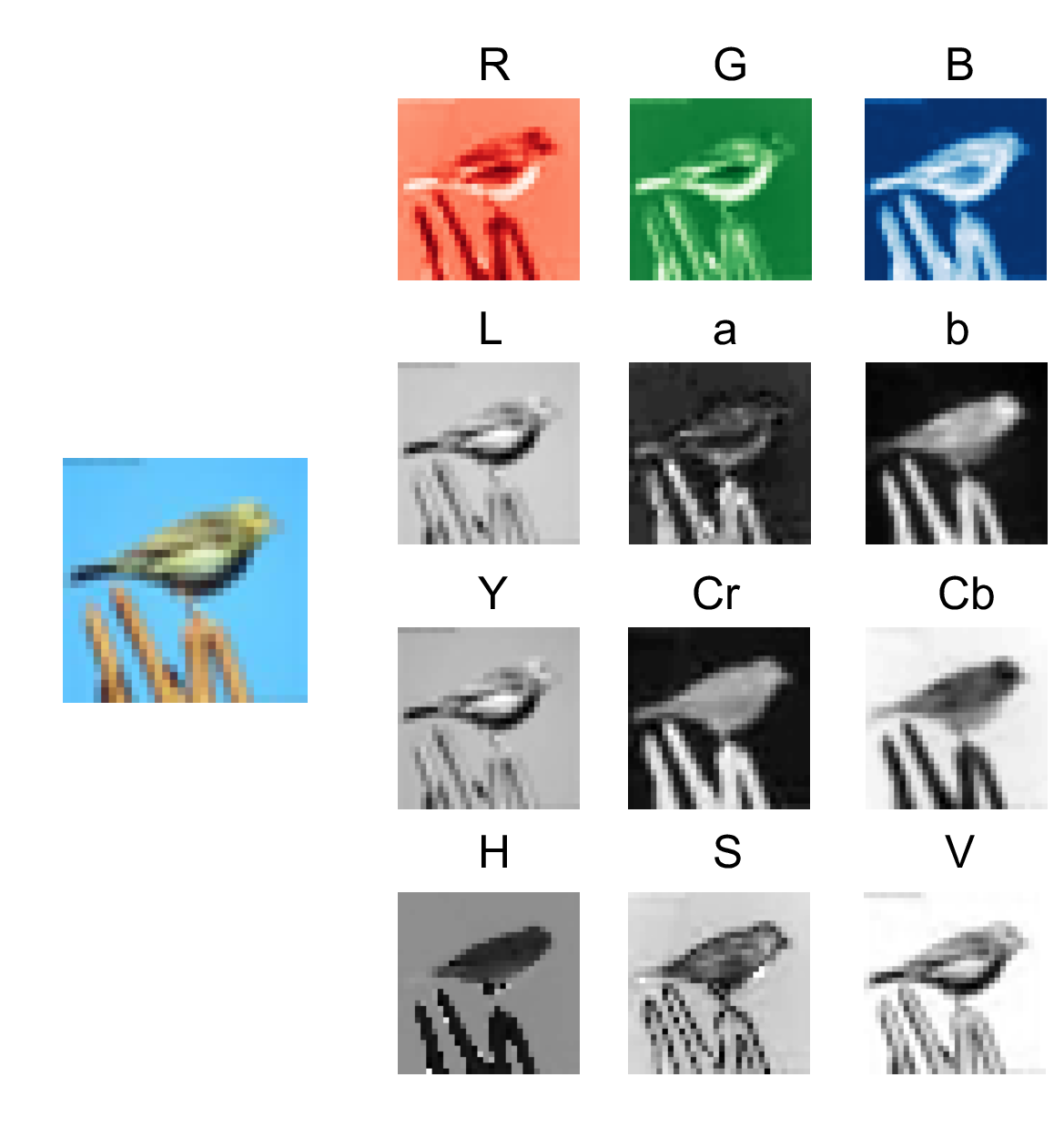}
   \caption{Plotted each channel of the image in four color spaces: RGB, Lab, YCrCb, and HSV.}
   \label{fig:images_full_color}
\end{figure}

Studies suggest that the selection of color space significantly affects the accuracy of classification model~\cite{23_1gowda2018colornet}.
Research on hybrid color spaces in CNNs has demonstrated improved accuracy, with a model utilizing 1.75M parameters and multiple color spaces performing comparably to one with 27M parameters. 
Converting images in different color spaces and visualizing their channel information, as shown in~\Cref{fig:images_full_color}, reveals noticeable differences in brightness and edge details, motivating the exploration of various color spaces to enhance model robustness and generalization.
Further extending QCNNs to these color spaces could provide valuable insights into performance across various color characteristics and highlight potential advantages in diverse image datasets.

\begin{figure*}[htbp]
  \centering
   \includegraphics[width=0.9\linewidth]{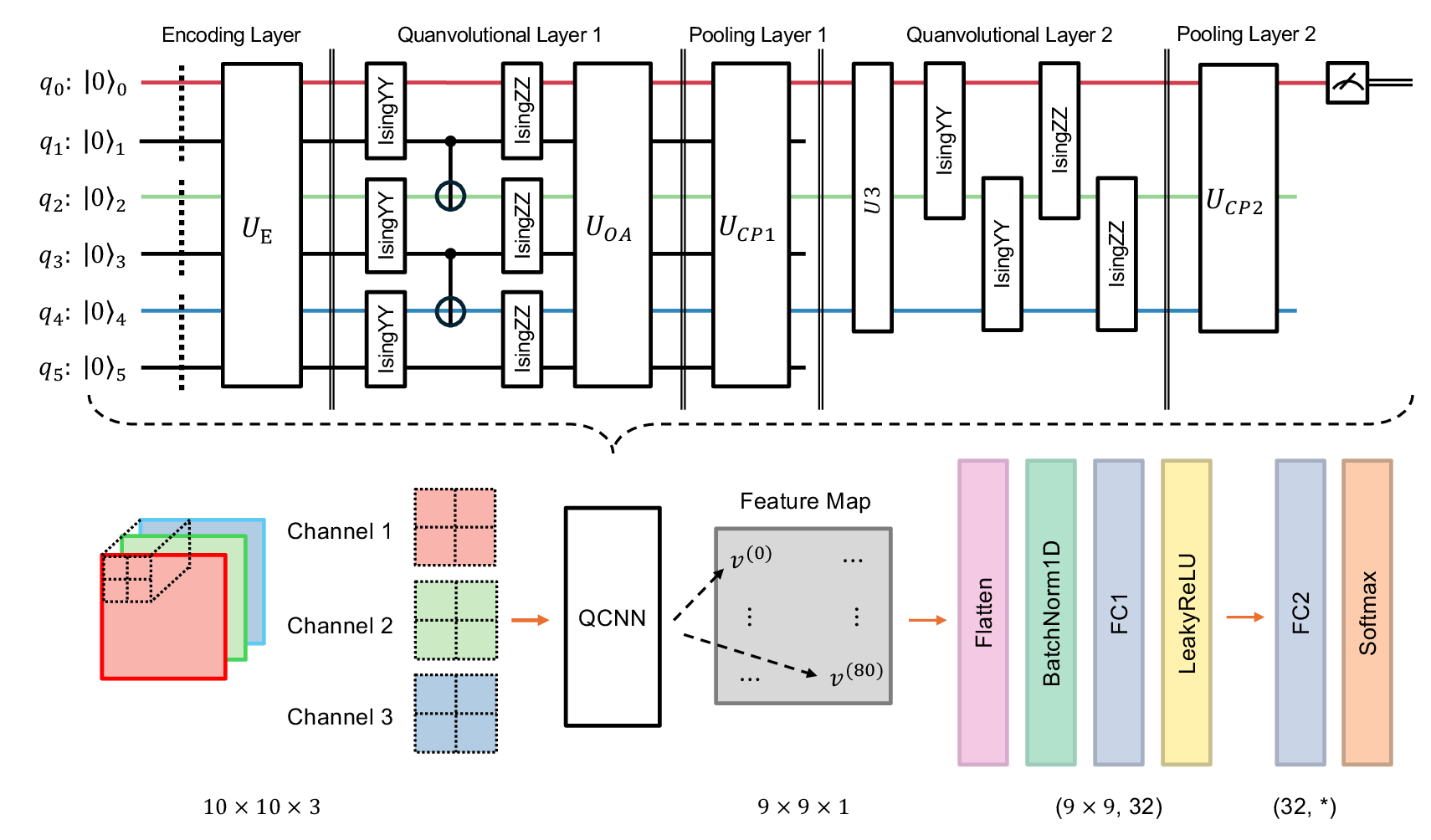}
   \caption{Details of our proposed HQCNN architecture, showing the main components of the QCNN. This includes the relationship between image channel encoding and qubit positions within the QCNN, and the output feature map size of the quantum layer.}
   \label{fig:img1_hqccnn}
\end{figure*}

To investigate the potential of QCNNs in various color spaces, we propose a hybrid quantum-classical convolutional neural network (HQCNN) model, the architecture of which is shown in~\Cref{fig:img1_hqccnn}. 
The HQCNN combines the strengths of quantum computing and classical neural networks: the quantum component extracts complex features, while the classical component handles fully connected layer operations and loss computation to optimize classification performance. 
Initially, the image dataset is scaled and converted into multiple color spaces, including Lab, YCrCb, and HSV. 
Each image is processed using a fixed $2 \times 2$ quantum convolutional kernel, encoding four pixel values from each channel into quantum states. After encoding, PQCs evolve the quantum states, followed by measurement to obtain feature values corresponding to the three channels.

The main contributions of this paper:

\textbf{(1)} This paper proposes a HQCNN model that is among the first to demonstrate image classification accuracy across various color spaces, offering a novel perspective on harnessing quantum advantages in computer vision. Given the limited exploration of this area in current quantum computing research, this study aims to contribute significantly to the field. For each color channel, after encoding and applying the first quanvolutional circuits, we introduce an optimized additional two-qubit parameterized unitary operator before the pooling layer to extract channel information and assess its impact on model performance.

\textbf{(2)} We analyze the types of quantum rotation gates used in the encoding layer and reduce the number of two-qubit gates in the quantum circuit, resulting in shallower circuits with fewer parameters. 
For each color channel, after encoding and applying the first quanvolutional circuits, we introduce an optimized additional two-qubit parameterized unitary operator before the pooling layer to extract channel feature and assess its impact on model performance. 
This flexibility facilitates exploration of quantum structures in specific tasks.

The remainder of this paper is organized as follows:~\Cref{sec:related} reviews current HQCNN applications in image classification. \Cref{sec:method} details the composition and implementation of the proposed HQCNN model. \Cref{sec:results} presents experimental results from classification tasks in various color spaces. Finally,~\Cref{sec:conclusion} summarizes the study and proposes future research directions.

\section{Related work}
\label{sec:related}
This section provides briefly overview of PQCs and reviews recent advancements in HQCNN models for color image classification.

The current era of quantum computing is characterized by noisy intermediate-scale quantum (NISQ) devices, where quantum hardware is prone to noise.
Consequently, designing quantum algorithms that function effectively on these noisy devices is essential. PQCs address this challenge by providing a flexible framework for quantum computing~\cite{24_benedetti2019parameterized}. 
Researchers have utilized PQCs to develop various quantum algorithms, such as the variational quantum eigensolver (VQE)~\cite{25_peruzzo2014variational} and the quantum approximate optimization algorithm (QAOA)~\cite{26_farhi2014quantum}. These algorithms have demonstrated potential in practical applications, including electronic structure research~\cite{27_kandala2017hardware} and solving combinatorial optimization problems, such as the MaxCut problem~\cite{28_wang2018quantum}.
The flexibility of PQC makes them an ideal choice for exploring potential quantum advantages in NISQ devices. It can approximate target values for specific tasks by adjusting the variational parameters. 
The QCNN model takes full advantage of this flexibility, showing promising performance in particular quantum computing tasks. Building on this foundation, HQCNN further integrates the robustness of classical algorithms, enhancing the reliability and stability of the final classification results.

Early work by Henderson et al.~\cite{40_henderson2020quanvolutional} compared the classification accuracy of their proposed QCNN with that of a classical CNN and a random model on the MNIST dataset. Their results demonstrated that, compared to the classical CNN, the accuracy of QCNN improved with more training iterations, and adding a certain number of quantum filters enhanced the model performance. 
In deeper hybrid models, quantum convolutional layers boosted network accuracy.
Ovalle-Magallanes et al.~\cite{29_ovalle2023quantum} conducted ablation studies on the QCNN model for the MNIST classification task, revealing that increasing the size and number of quanvolutional kernels in the PQC does not always result in optimal classification accuracy.
Liu et al~\cite{30_liu2021hybrid} proposed an early HQCNN for image classification tasks. This model uses a global phase measurement method to measure all qubits and generate multi-channel feature maps. The classification results on the Tetris dataset have shown that it could even surpass the accuracy of classical CNN.

Jing et al.~\cite{32_jing2022rgb} proposed two quantum convolutional circuits that compare different methods for extracting inter-channel and intra-channel information. Experimental results show that these newly proposed quantum convolutional circuits achieve higher test accuracy than classical CNN in RGB image classification tasks. Moreover, their work demonstrates that the design of larger PQCs can improve the performance of multi-class classification tasks.

Riaz et al.~\cite{33_riaz2023accurate} proposed a QCNN requiring no parameter updates, which showed outperforming QCNN with randomly generated PQCs on the MNIST and CIFAR-10 datasets. However, its accuracy decreased on the German Traffic Sign Recognition Benchmark (GTSRB) dataset, suggesting that the application of quantum algorithm-based classifiers requires further investigation for colored and complex datasets. Smaldone et al.~\cite {34_smaldone2023quantum} developed a QCNN for multi-channel image processing, which achieved higher accuracy on the CIFAR-10 dataset compared to prior work~\cite{32_jing2022rgb}, with reduced quantum circuit complexity.

Yang and Sun~\cite{41_yang2022semiconductor} proposed a hybrid quantum-classical model for semiconductor wafer defect detection, conducted an ablation study on various PQC~\cite{43_sim2019expressibility}. The experimental results showed that angle encoding outperformed basis state and amplitude encoding, with stronger entangled and expressive circuits showing clear advantages. 
Their hybrid model achieved higher accuracy with limited quantum resources than classical CNN on the LIT Hotspot (ICCAD-2012) and Defect Wafer Map (WM-811K) datasets.
Senokosov et al.~\cite{31_senokosov2024quantum} proposed two hybrid neural networks. The first model, using parallelized quantum circuits, achieved over $99\%$ accuracy on MNIST and Medical MNIST and over $82\%$ on CIFAR-10 with eight times fewer parameters than classical models. The second model featuring quantum convolutional layers achieved performance comparable to classical CNNs while using four times fewer parameters.

Research on QCNN-based color image classification has predominantly focused on the RGB color spaces, while broader studies have mainly addressed grayscale image processing~\cite{39_kharsa2023advances}. However, the application of QCNNs or hybrid quantum models to image classification tasks in alternative color spaces remains limited.

\section{Methods}
\label{sec:method}
The HQCNN model proposed in this study integrates quantum and classical components to utilize the strengths of both paradigms. 
The quantum component encodes classical image data into quantum states, utilizing the high-dimensional Hilbert spaces of quantum computing. These states evolve through PQC, and local measurements are performed on qubits to extract image features while mitigating vanishing gradients associated with the barren plateau problem~\cite{19_cerezo2021cost}.
This approach enables direct quantum processing of raw input data without requiring additional preprocessing through a classical CNN.

\subsection{Quantum component}
The quantum component consists of three primary layers: encoding, quanvolutional, and pooling layers, executed sequentially, following the quantum circuit structure proposed by Barenco et al.~\cite{42_barenco1995elementary}.

\subsubsection{Encoding layer}
In the quantum encoding (embedding) layer, single-qubit parameterized rotation gates $R_x(\theta)$, $R_y(\theta)$, and $R_z(\theta)$ are employed for encoding.

\begin{equation}
    R_x(\theta) = e^{-i \theta \sigma_x / 2} = \begin{bmatrix}
        \cos(\theta/2) & -i \sin(\theta/2) \\
        -i \sin(\theta/2) & \cos(\theta/2)
    \end{bmatrix}
    \label{eq:rotation_x}
\end{equation}

\begin{equation}
    R_y(\theta) = e^{-i \theta \sigma_y / 2} = \begin{bmatrix}
    \cos(\theta/2) & -\sin(\theta/2) \\
    \sin(\theta/2) & \cos(\theta/2)
    \end{bmatrix}
    \label{eq:rotation_y}
\end{equation}

\begin{equation}
    R_z(\theta) = e^{-i \theta \sigma_z / 2} = \begin{bmatrix}
    e^{-i\theta/2} & 0 \\
    0 & e^{i\theta/2}
    \end{bmatrix}
    \label{eq:rotation_z}
\end{equation}

\begin{figure}[htbp]
  \centering
   \includegraphics[width=1\linewidth]{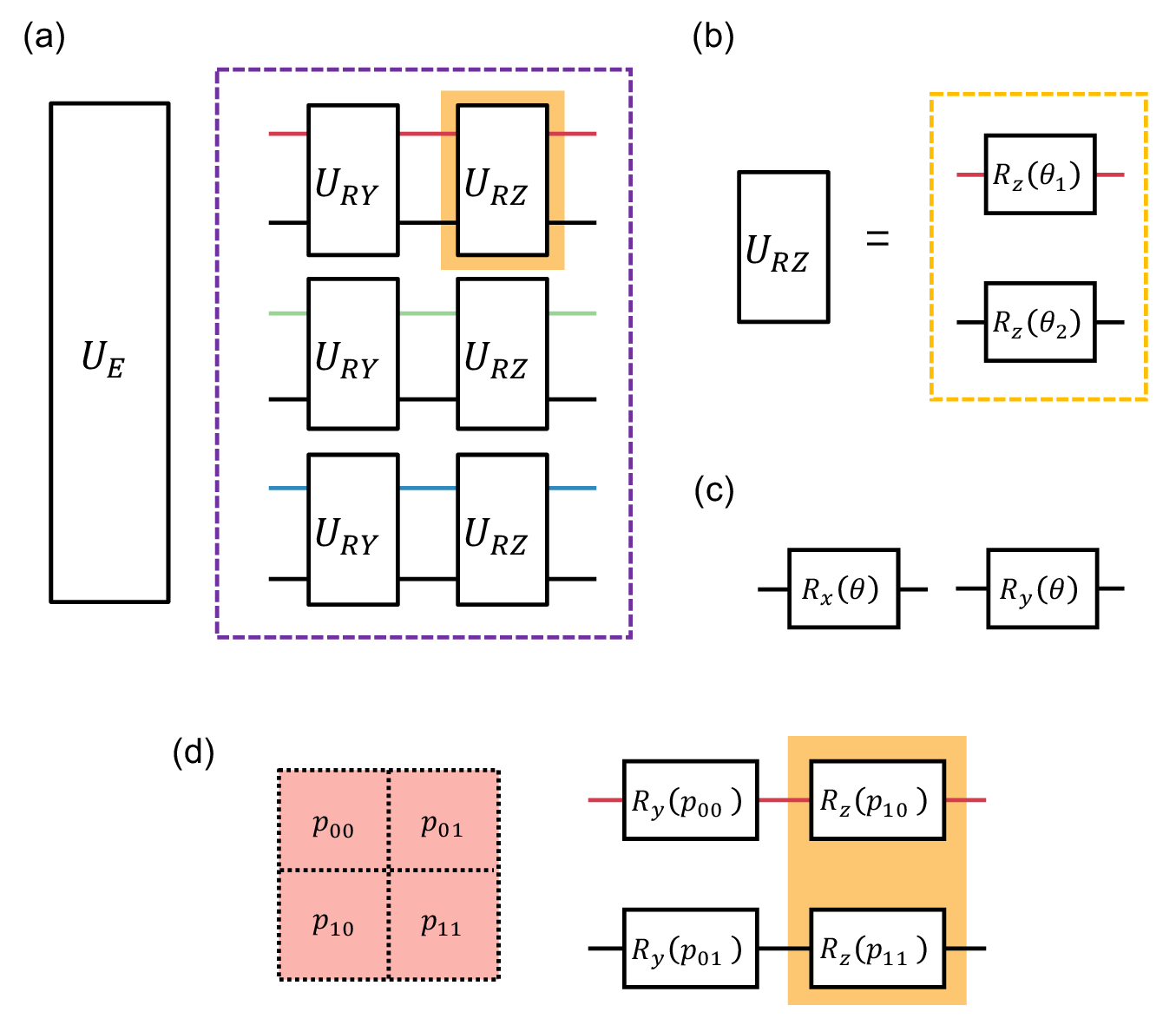}
   \caption{
       (a) The structure of the quantum encoding layer in QCNN, exemplifying the $R_{YZ}$ rotation gate combination (along with $R_{XZ}$ and $R_{XY}$ combinations).
       (b) The structure of the $U_{RZ}$ unitary operator module, applying an $R_z(\theta)$ operator on each circuit.
       (c) The representation of $R_x(\theta)$ and $R_y(\theta)$ operators in the quantum circuit.
       (d) An example of selecting the R channel of an RGB image using a $2 \times 2$ kernel, where the four pixels are sequentially encoded into corresponding rotation gates at specific positions.
   }
   \label{fig:img2_encoding}
\end{figure}

The structure of the unitary operators $U_{RY}$ and $U_{RZ}$ in the quantum circuit, as depicted in~\Cref{fig:img2_encoding}, utilizes two distinct single-qubit rotation gates per qubit to provide local degrees of freedom, preparing the quantum states of the image pixels.

For the R channel of an RGB image, the quanvolutional kernel extracts four pixel values from the channel. These values are encoded into quantum states as follows: the first two values are processed using angle encoding via the $R_y(\theta)$ gate, while the remaining two are encoded with the $R_z(\theta)$ gate, as depicted in~\Cref{fig:img2_encoding} (d). This encoding process is applied to each of the three channels, resulting in twelve encoded pixel values for the selected kernel. Each set of twelve pixel values corresponds to the image pixels selected by the quanvolution kernel in three channels.

\subsubsection{Quanvolutional layer}
\begin{figure}[htb]
  \centering
   \includegraphics[width=1.05\linewidth]{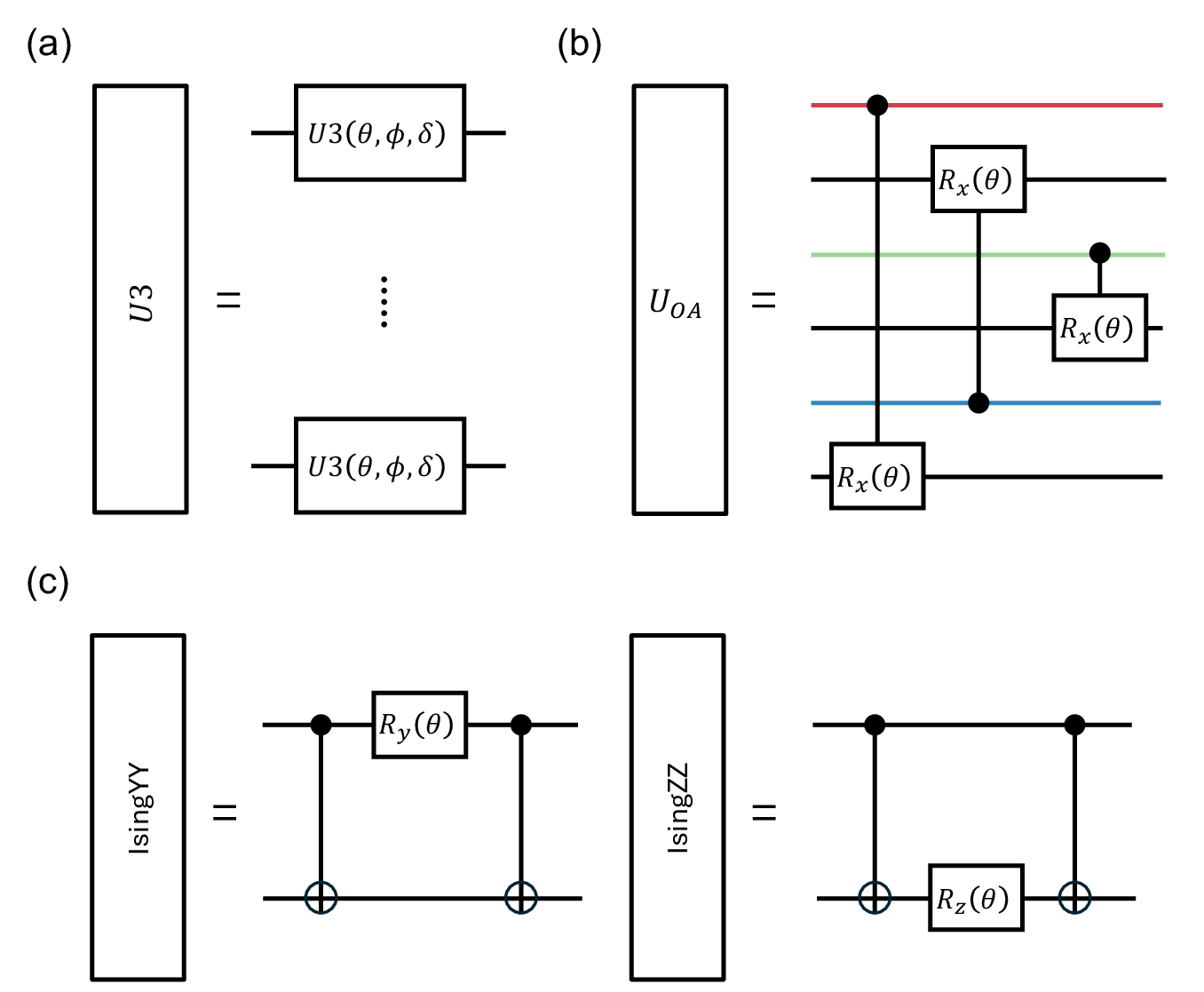}
   \caption{
       (a) The unitary operator $U3$ consists of independently applying a single-qubit parameterized gate $U3(\theta, \phi, \delta)$ to each qubit.
       (b) An example of the composition of the optimized additional unitary operator module $U_{OA}$, utilizing type of $CR_x(\theta)$ gate, with optional controlled $R_y(\theta)$ and $R_z(\theta)$ gates.
       (c) The structure of the IsingYY and IsingZZ unitary operators in the quantum circuit.
   }
   \label{fig:img3_U3_UOA}
\end{figure}

Following encoding, the first quanvolutional layer employs two distinct parameterized two-qubit operators, IsingYY and IsingZZ, each implemented using two CNOT gates for entanglement and a parameterized rotation gate, as shown in~\Cref{fig:img3_U3_UOA} (c).
For an RGB image, as an illustrative example, the four pixel values of each color channel are mapped to two qubits via angle encoding, with a total of six qubits representing the red, blue, and green channels.

Initially, the IsingYY unitary operator is applied to the two qubits within each channel, using trainable parameters to provide non-local degrees of freedom. This generates and optimizes entangled states, enhancing the expressiveness of the variational quantum circuit (VQC), also referred to as the PQC, for complex data.
Subsequently, CNOT gates are applied to create cross-channel entanglement, specifically between the R and G channels ($q_1$-control, $q_2$-target) and the G and B channels ($q_3$-control, $q_4$-target), resulting in a globally entangled state across all six qubits.
Finally, the IsingZZ unitary operator is similarly applied. It introduces relative phases through trainable parameters to optimize the phase structure of the entangled state, further improving the PQC's expressiveness for image data.

\begin{equation}
    CNOT = \begin{bmatrix}
    1 & 0 & 0 & 0 \\
    0 & 1 & 0 & 0 \\
    0 & 0 & 0 & 1 \\
    0 & 0 & 1 & 0
    \end{bmatrix}
    \label{eq:cnot_matrix}
\end{equation}

To further enhance the model, an additional optimized unitary operator module $U_{OA}$ is introduced between the first quanvolutional and pooling layers, as shown in~\Cref{fig:img3_U3_UOA} (b).
The $U_{OA}$ operator comprises three parameterized controlled rotation gates, which adjust the entanglement structure across channels to capture non-local correlations of pixel features, enabling the evaluation of the impact of different parameterized gates on the expressiveness of the QCNN.

\begin{equation}
    CR_x(\theta) = \begin{bmatrix}
    1 & 0 & 0 & 0 \\
    0 & 1 & 0 & 0 \\
    0 & 0 & \cos(\theta/2) & -i \sin(\theta/2) \\
    0 & 0 & -i \sin(\theta/2) & \cos(\theta/2)
    \end{bmatrix}
    \label{eq:controlled_rotation_x}
\end{equation}

Sim et al.~\cite{43_sim2019expressibility} found that the $CR_x(\theta)$ gate outperforms the $CR_z(\theta)$ gate in expressiveness and entanglement capability. Accordingly, ablation experiments are conducted using three types of parameterized two-qubit gates, $CR_x(\theta)$, $CR_y(\theta)$, and $CR_z(\theta)$, to assess their effects.

\begin{equation}
    U3(\theta, \phi, \delta) = \begin{bmatrix}
        \cos(\theta/2) & -e^{i \delta} \sin(\theta/2) \\
        e^{i \phi} \sin(\theta/2) & e^{i (\phi + \delta)} \cos(\theta/2)
    \end{bmatrix}
    \label{eq:u3}
\end{equation}

Following the first pooling layer, the second quanvolutional layer, as shown in~\Cref{fig:img1_hqccnn}, applies the parameterized $U3$ gate to qubits $q_0$, $q_2$, and $q_4$, using trainable parameters $(\theta, \phi, \delta)$ to adjust local amplitudes and phases, thereby improving the expressiveness of the main features.
Subsequently, IsingYY and IsingZZ unitary operators are applied to adjust interactions between color channels, strengthening non-local feature correlations among the main feature qubits across the three channels.

\subsubsection{Pooling layer}

\begin{figure}[htb]
  \centering
   \includegraphics[width=1\linewidth]{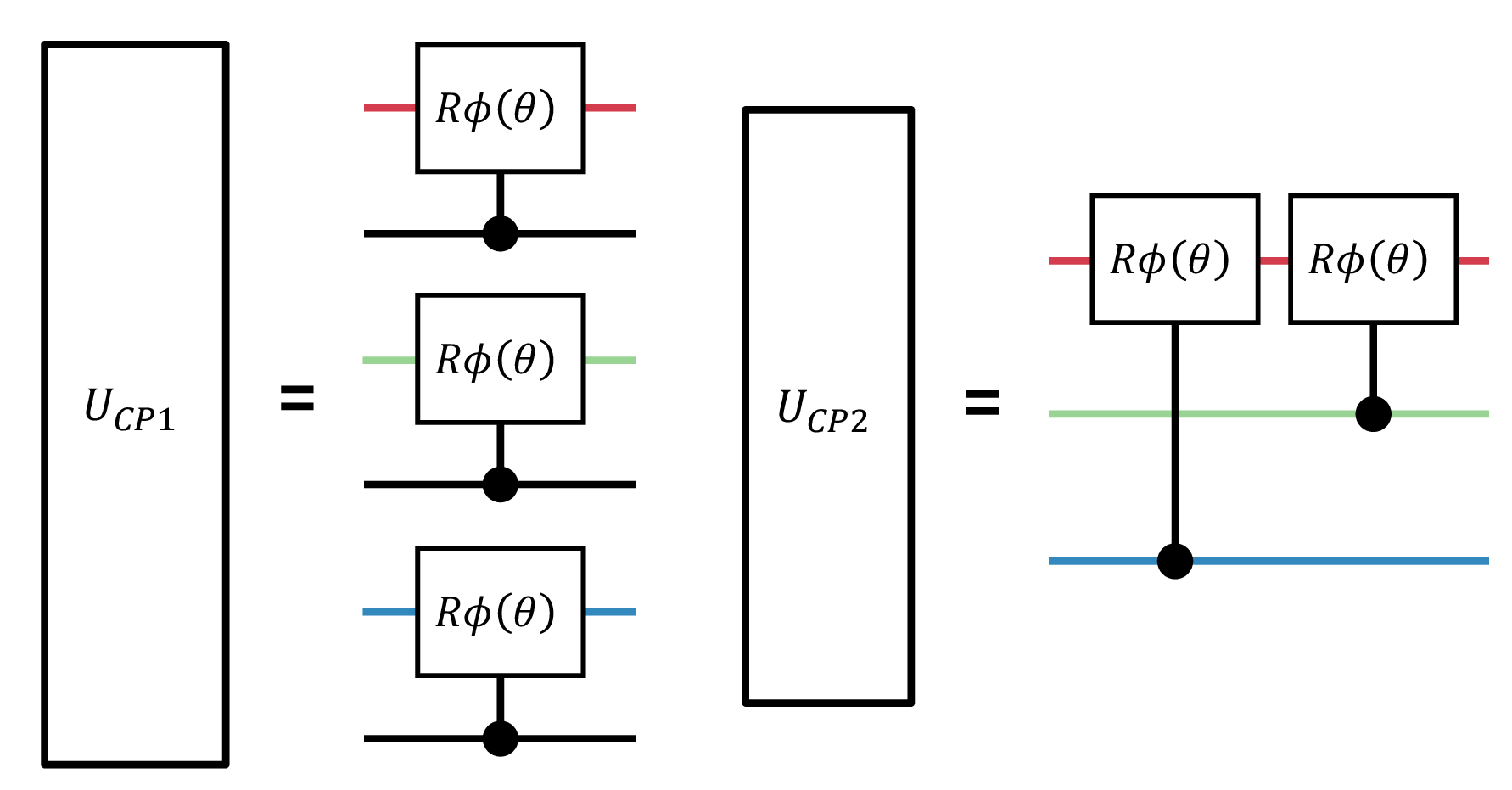}
   \caption{
       The structure of the $U_{CP1}$ and $U_{CP2}$ unitary operators, each comprising a distinct quantum pooling layer implemented with $CP(\theta)$.
   }
   \label{fig:img4_UCP}
\end{figure}

The first pooling layer employs the unitary operator $U_{CP1}$, as shown in~\Cref{fig:img4_UCP} (a), consisting of three controlled-phase gates $CP(\theta)$.
These gates utilize trainable parameters to provide non-local degrees of freedom, adjusting the phase structure across the red, green, and blue channels.
Specifically, $q_0$, $q_2$, and $q_4$ serve as target qubits, representing the main features of each channel and are retained for subsequent processing, while other qubits are disregarded to emulate classical pooling functionality.

The second pooling layer uses the unitary operator $U_{CP2}$, as shown in~\Cref{fig:img4_UCP} (b), where two parameterized controlled gates compress the features of the blue and green channels into the red channel.

\begin{equation}
    CP(\theta)=CR_{\phi}(\theta) = \begin{bmatrix}
    1 & 0 & 0 & 0 \\
    0 & 1 & 0 & 0 \\
    0 & 0 & 1 & 0 \\
    0 & 0 & 0 & e^{i\theta}
    \end{bmatrix}
    \label{eq:conctrolled_phase}
\end{equation}

\subsubsection{Measurement}

The quantum state is processed by a parameterized unitary $U_{Ansatz}(\theta)$ and an angle encoding unitary $U_E(\theta)$, acting on six qubits. The circuit consists of angle encoding, two quanvolutional layers, and two pooling layers. Finally, the Pauli-Z expectation value is computed for qubit $q_0$, resulting in

\begin{equation}
    v^{(i)}=\langle 0|U_{E}^\dagger(x_i) U_{Ansatz}(\theta)^\dagger Z^{0} U_{Ansatz}(\theta) U_{E}(x_i) |0\rangle,
    \label{eq:expvalue}
\end{equation}

where $Z^{0}$ denotes the Pauli-Z operator on $q_0$, $U_{Ansatz}(\theta)$ encompasses all quanvolutional and pooling operations, and $x$ denotes the 12 pixels extracted by the quanvolutional kernel, with $i$ indicating the index, corresponding to the feature map ($v^{(i)}$) as shown in~\Cref{fig:img1_hqccnn}.
The expectation value $v^{(i)}$ is calculated using the reduced density matrix of $q_0$,
\begin{equation}
    \rho_0 = Tr_{1,2,3,4,5}(|\psi_i \rangle \langle \psi_i|)
    \label{eq:rdm}
\end{equation}
with $|\psi_i\rangle = U_{Ansatz}(\theta)U_E(x_i) |0\rangle^{\otimes 6}$, thus $v^{(i)} = Tr_i(\rho_0 Z)$.
This expectation value represents the feature encoded in $q_0$, serving as the output for cross-channel pixel characteristics in quantum image processing tasks.

\subsection{Classical component}
The classical component of the HQCNN model primarily consists of fully connected layers and classical optimization algorithms.

After processing through the QCNN, each image generates a feature map derived from the Pauli-Z expectation values of the output qubits, as depicted in~\Cref{fig:img1_hqccnn}. This feature map is flattened into a one-dimensional vector, normalized by a BatchNorm1D layer, and passed to the first fully connected classical layer (FC1).
A leaky rectified linear unit (LeakyReLU) activation function~\cite{35_maas2013rectifier} is applied to ensure effective backpropagation, even for negative inputs. The output is then fed into the second fully connected layer (FC2), with its output size matching the number of classes in the classification task. A softmax function~\cite{36_williams1989learning} converts the FC2 outputs into a probability distribution. Finally, the cross-entropy loss function~\cite{37_rumelhart1986learning} computes the error between predicted and true labels, enabling backpropagation to update the parameters of both quantum and classical components.

\begin{figure}[htb]
  \centering
  \includegraphics[width=0.7\linewidth]{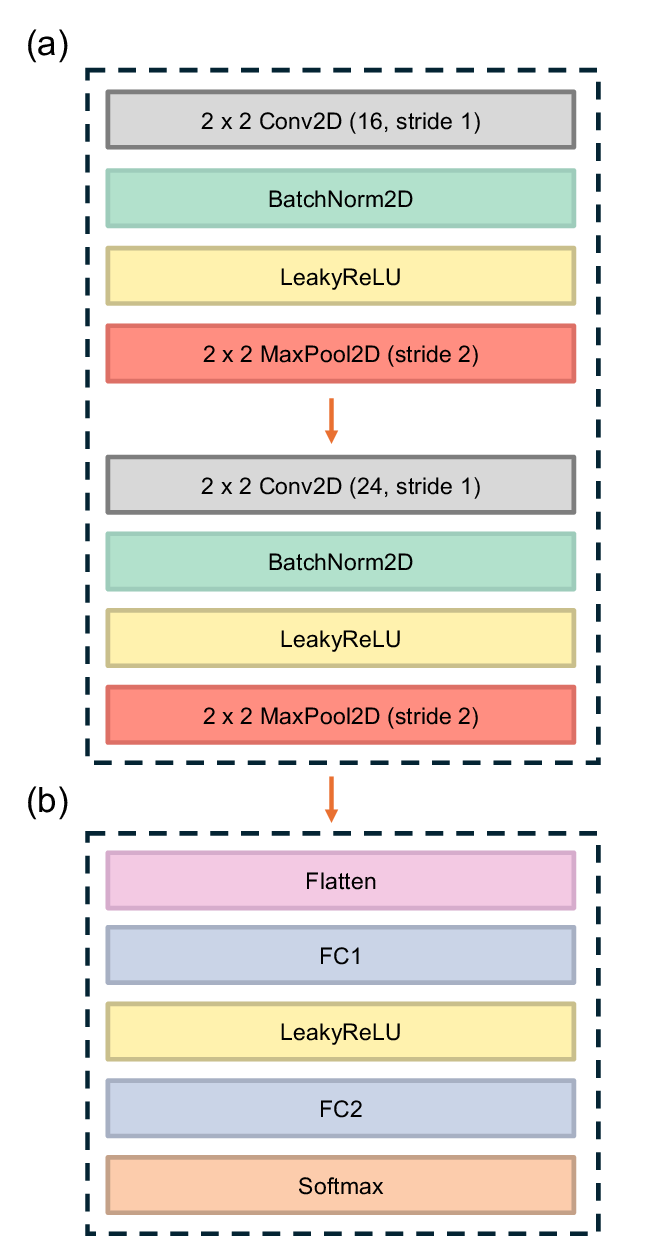}
   \caption{
       (a) Two convolutional and pooling layers. (b) Similar to the classical part of the HQCNN, omitting the BatchNorm1D layer.
   }
   \label{fig:cnn}
\end{figure}

\subsection{Simple CNN}
To evaluate the performance of the HQCNN model, a simple classical CNN model is introduced as a baseline for comparison. This classical model adopts a structure similar to that of the HQCNN, as shown in~\Cref{fig:cnn}.

\subsection{Model scale comparison}
The quantum circuit complexity of the HQCNN is analyzed using the gate set $\mathcal{G} = \{CNOT, R_x, R_y, R_z\}$ as we consider. The QCNN decomposition results in 48 CNOT gates, a circuit depth of 20, and 27 trainable parameters.
For binary classification tasks, the classical component of the HQCNN includes 2852 trainable parameters, compared to 2762 parameters for the simple CNN. In the ten-class classification task, only the output dimension of the FC2 differs, adding 264 parameters to the binary classification baseline. Consequently, the HQCNN has 117 more parameters than the classical CNN.

\section{Experiments}
\label{sec:results}

\subsection{Datasets}
To evaluate the performance of the HQCNN and the classical CNN, four datasets are selected: CIFAR-10~\cite{44_krizhevsky2009learning}, EuroSAT~\cite{45_helber2019eurosat}, and SAT-4~\cite{46_basu2015deepsat} for binary classification tasks, and MNIST~\cite{47_deng2012mnist} for a ten-class classification task.
CIFAR-10 comprises complex color images that challenge the expressive capabilities of models.
EuroSAT and SAT-4, both satellite image datasets, assess the generalization in domain-specific tasks.
MNIST, due to its simplicity and widespread use, serves as a benchmark for evaluating basic deep learning models.
Compared to the large-scale ImageNet dataset, these smaller datasets are better suited for evaluating lightweight HQCNN and CNN architectures, which face challenges with ImageNet’s computational and data complexity.
The single-channel grayscale images of MNIST are transformed into three-channel images by replicating the grayscale channel to align with the three-channel format of color datasets.

All datasets are preprocessed consistently, with images resized to $(10 \times 10 \times 3)$ ($\mathcal{H} \times \mathcal{W} \times \mathcal{C}$) using bilinear interpolation, and RGB images are converted into Lab, YCrCb, and HSV color spaces using OpenCV~\cite{48_bradski2000opencv}.
To address varying color space ranges, the L channel of Lab, the Y channel of YCrCb, HSV, and RGB are normalized to $[0, 1]$, while the a and b channels of Lab and the Cr and Cb channels of YCrCb are normalized to $[-1, 1]$.
Each dataset is randomly sampled without replacement and split into a training set of 500 images, a validation set of 100 images, and a test set of 100 images.

\subsection{Experimental setup}
A quantum convolutional kernel of size $2 \times 2$ is applied across all three channels of the input image with a stride of $1$, producing an output of size $2 \times 2 \times 3$.
The HQCNN and CNN models are trained using the AdamW optimizer~\cite{49_loshchilov2017decoupled} to update parameters, with the ReduceLROnPlateau scheduler employed for dynamic learning rate adjustment.
The initial learning rate is set to $0.01$, reduced to $10\%$ of its current value (with a minimum of $10^{-4}$) if the validation loss does not decrease for three consecutive epochs over a total of $20$ epochs.
In the classical component of both models, the FC1 has an output size of $32$, while the FC2 matches the number of classes in the classification task.
The QCNN is constructed and simulated using the PennyLane framework~\cite{50_bergholm2018pennylane}, while the HQCNN and CNN models are implemented and trained using the PyTorch library~\cite{51_paszke2019pytorch}.

Parameters of both quantum and classical components are optimized via the backpropagation algorithm, implemented in PyTorch to compute gradients of the cross-entropy loss function, enabling optimization through gradient descent.
In HQCNN simulation, the backpropagation method provides faster gradient computation compared to the parameter shift rule~\cite{52_wierichs2022general}, which is typically required for quantum hardware.
Due to hardware limitations, direct application of backpropagation is not feasible on quantum hardware, necessitating the use of the parameter shift rule for gradient computation. 
All quantum simulations and computations for both HQCNN and classical CNN models on an Apple M4 CPU.

\subsection{Evaluation metrics}
For both binary and ten-class classification tasks, the performance of the HQCNN and classical CNN models is assessed using macro-averaged precision, recall, F1 score, and accuracy. 
These metrics are computed as arithmetic averages across all classes to account for balanced class distributions in the selected datasets. The definitions of the metrics are as follows:

\begin{itemize}
    \item True Positives ($TP_i$): The number of correctly predicted instances for class $i$.
    \item False Positives ($FP_i$): The number of instances incorrectly predicted as class $i$.
    \item False Negatives ($FN_i$): The number of instances of class $i$ incorrectly predicted as another class.
\end{itemize}

The precision ($P_i$) and recall ($R_i$) for each class $i$ are defined as:
\begin{equation}
    P_i = \frac{TP_i}{TP_i + FP_i},
    \label{eq:precision}
\end{equation}
\begin{equation}
    R_i = \frac{TP_i}{TP_i + FN_i}.
    \label{eq:recall}
\end{equation}

Given the balanced class sizes in the datasets, accuracy is calculated as the proportion of correctly classified instances across all samples, where ${N}$ denotes the total number of samples:

\begin{equation}
    Accuracy = \frac{1}{N}\sum_{i=0}^{N-1}TP_i.
    \label{eq:accuracy}
\end{equation}

Macro-averaged metrics are computed as follows, where $n$ represents the number of classes:
\begin{equation}
    Macro-Precision = \frac{1}{n}\sum_{i=1}^{n} P_i,
    \label{eq:macro-precision}
\end{equation}
\begin{equation}
    Macro-Recall = \frac{1}{n}\sum_{i=1}^{n} R_i,
    \label{eq:macro-recall}
\end{equation}
\begin{equation}
    Macro-F1 = \frac{1}{n}\sum_{i=1}^{n}2\cdot \frac{P_i\cdot R_i}{P_i + R_i}.
    \label{eq:macro-f1}
\end{equation}
These metrics provide a comprehensive evaluation of model performance across different classification tasks.

\subsection{CIFAR-10}
Two groups of classes, $\textrm{C}_1$ (airplane-frog) and $\textrm{C}_2$ (frog-ship) are selected from the CIFAR-10 dataset to evaluate the performance of the CNN and HQCNN models. 
As shown in~\Cref{tab:1_cifar10}, the CNN model achieves an accuracy of $95\%$ in the YCrCb on the $\textrm{C}_1$ test set, outperforming other color spaces. 
For $\textrm{C}_2$, accuracies of $96\%$ are obtained in both Lab and YCrCb, surpassing the RGB performance of $94\%$. 
In the HQCNN model, the default encoding employs the $R_{XZ}$ operator. 
When the $U_{OA}$ operator employs the $R_x$ gate, accuracies of $94\%$ are achieved in both RGB and Lab for $\textrm{C}_1$, outperforming configurations using $R_y$ and $R_z$ gates. 
For $\textrm{C}_2$, the HQCNN in HSV achieves the highest accuracy of $97\%$, compared to $94.5\%$ in RGB and the CNN accuracies across all four color spaces.

Average accuracy and standard deviation from five training runs, as shown in~\Cref{tab:2_cifar10_mean}.
For the CNN, an average accuracy of $94.1\%$ is achieved in the YCrCb color space for the $\textrm{C}_1$ task, outperforming RGB $91.7\%$. 
In $\textrm{C}_2$ task, the CNN achieves an accuracy of $94.5\%$ in Lab and YCrCb, while RGB reaches $93.3\%$ and HSV $92.5\%$. 
For the HQCNN, no encoding combination surpasses the highest CNN accuracy.
However, the $U_{OA}$ operator with the $R_x$ gate ($U_E: R_{XZ}$) in the Lab color space achieves an average accuracy of $93\%$ for $\textrm{C}_1$, outperforming HQCNN’s best RGB performance of $90.3\%$ ($R_{XY}-R_y$). 
Similar accuracies are obtained with the $R_x$ and $R_y$ gates, while the $R_z$ gate, though generally less effective, remains competitive in HSV.

\begin{table*}[ht]
    \caption{The best performance metrics results ($\%$) of the CNN and HQCNN models on the CIFAR-10 dataset for two binary classification tasks.}
    \begin{adjustbox}{center}
    \centering
    \footnotesize
    \begin{tabular}{c * {11}{l}}
        \toprule
        \multicolumn{1}{l}{\multirow{1}{*}{Model}} &
        \multicolumn{2}{l}{\multirow{1}{*}{$U_{OA}$}} &
        \multicolumn{2}{l}{\multirow{1}{*}{Dataset}} &
        \multicolumn{4}{c}{\multirow{1}{*}{CIFAR-10}}
        \\
        \cmidrule(lr){4-11}
        & & & \multicolumn{1}{l}{$\mathrm{C}_{1}^\mathrm{RGB\ \ }$} & \multicolumn{1}{l}{$\mathrm{C}_{1}^\mathrm{Lab\ \ }$} &
        \multicolumn{1}{l}{$\mathrm{C}_{1}^\mathrm{YCrCb}$} & \multicolumn{1}{l}{$\mathrm{C}_{1}^\mathrm{HSV\ \ }$} &
        \multicolumn{1}{l}{$\mathrm{C}_{2}^\mathrm{RGB\ \ }$} & \multicolumn{1}{l}{$\mathrm{C}_{2}^\mathrm{Lab\ \ }$} &
        \multicolumn{1}{l}{$\mathrm{C}_{2}^\mathrm{YCrCb}$} & \multicolumn{1}{l}{$\mathrm{C}_{2}^\mathrm{HSV\ \ }$}
        \\
        \midrule
        \multicolumn{1}{l}{\multirow{4}{*}{{CNN}}} &
        \multicolumn{1}{l}{\multirow{4}{*}{ }} &
        \multicolumn{1}{l}{Acc.} &
        \multicolumn{1}{l}{$93.5$} & \multicolumn{1}{l}{$94.5$} &
        \multicolumn{1}{l}{$\textbf{95.0}$} & \multicolumn{1}{l}{$92.0$} &
        \multicolumn{1}{l}{$94.0$} & \multicolumn{1}{l}{$\textbf{96.0}$} &
        \multicolumn{1}{l}{$\textbf{96.0}$} & \multicolumn{1}{l}{$94.5$}
        \\
        & & \multicolumn{1}{l}{F1} &
        \multicolumn{1}{l}{$93.4$} & \multicolumn{1}{l}{$94.4$} &
        \multicolumn{1}{l}{$95.0$} & \multicolumn{1}{l}{$91.9$} &
        \multicolumn{1}{l}{$93.9$} & \multicolumn{1}{l}{$95.9$} &
        \multicolumn{1}{l}{$96.0$} & \multicolumn{1}{l}{$94.4$}
        \\
        & & \multicolumn{1}{l}{Rec.} &
        \multicolumn{1}{l}{$93.5$} & \multicolumn{1}{l}{$94.5$} &
        \multicolumn{1}{l}{$95.0$} & \multicolumn{1}{l}{$92.0$} &
        \multicolumn{1}{l}{$94.0$} & \multicolumn{1}{l}{$96.0$} &
        \multicolumn{1}{l}{$96.0$} & \multicolumn{1}{l}{$94.5$}
        \\
        & & \multicolumn{1}{l}{Pre.} &
        \multicolumn{1}{l}{$93.5$} & \multicolumn{1}{l}{$94.5$} &
        \multicolumn{1}{l}{$95.0$} & \multicolumn{1}{l}{$92.1$} &
        \multicolumn{1}{l}{$94.0$} & \multicolumn{1}{l}{$96.0$} &
        \multicolumn{1}{l}{$96.0$} & \multicolumn{1}{l}{$94.5$}

        \\
        \cmidrule(lr){1-11}
        \multicolumn{1}{l}{\multirow{12}{*}{HQCNN}} &
        \multicolumn{1}{c}{\multirow{4}{*}{$R_x$}} &
        \multicolumn{1}{l}{Acc.} &
        \multicolumn{1}{l}{$\textbf{94.0}$} & \multicolumn{1}{l}{$\textbf{94.0}$} &
        \multicolumn{1}{l}{$92.5$} & \multicolumn{1}{l}{$93.0$} &
        \multicolumn{1}{l}{$94.5$} & \multicolumn{1}{l}{$94.0$} &
        \multicolumn{1}{l}{$95.0$} & \multicolumn{1}{l}{$\textbf{97.0}$}
        \\
        & & \multicolumn{1}{l}{F1} &
        \multicolumn{1}{l}{$93.9$} & \multicolumn{1}{l}{$93.9$} &
        \multicolumn{1}{l}{$92.4$} & \multicolumn{1}{l}{$92.9$} &
        \multicolumn{1}{l}{$94.4$} & \multicolumn{1}{l}{$93.9$} &
        \multicolumn{1}{l}{$94.9$} & \multicolumn{1}{l}{$96.9$}
        \\
        & & \multicolumn{1}{l}{Rec.} &
        \multicolumn{1}{l}{$94.0$} & \multicolumn{1}{l}{$94.0$} &
        \multicolumn{1}{l}{$92.5$} & \multicolumn{1}{l}{$93.0$} &
        \multicolumn{1}{l}{$94.5$} & \multicolumn{1}{l}{$94.0$} &
        \multicolumn{1}{l}{$95.0$} & \multicolumn{1}{l}{$97.0$}
        \\
        & & \multicolumn{1}{l}{Pre.} &
        \multicolumn{1}{l}{$94.1$} & \multicolumn{1}{l}{$94.1$} &
        \multicolumn{1}{l}{$92.8$} & \multicolumn{1}{l}{$93.0$} &
        \multicolumn{1}{l}{$94.8$} & \multicolumn{1}{l}{$94.1$} &
        \multicolumn{1}{l}{$95.1$} & \multicolumn{1}{l}{$97.0$}
        \\
        \cmidrule(lr){2-11}
        \multicolumn{1}{l}{ } &
        \multicolumn{1}{c}{\multirow{4}{*}{$R_y$}} &
        \multicolumn{1}{l}{Acc.} &
        \multicolumn{1}{l}{$91.0$} & \multicolumn{1}{l}{$91.5$} &
        \multicolumn{1}{l}{$\textbf{93.0}$} & \multicolumn{1}{l}{$91.0$} &
        \multicolumn{1}{l}{$94.5$} & \multicolumn{1}{l}{$\textbf{95.5}$} &
        \multicolumn{1}{l}{$95.0$} & \multicolumn{1}{l}{$93.0$}
        \\
        & & \multicolumn{1}{l}{F1} &
        \multicolumn{1}{l}{$90.9$} & \multicolumn{1}{l}{$91.4$} &
        \multicolumn{1}{l}{$92.9$} & \multicolumn{1}{l}{$90.9$} &
        \multicolumn{1}{l}{$94.4$} & \multicolumn{1}{l}{$95.4$} &
        \multicolumn{1}{l}{$94.9$} & \multicolumn{1}{l}{$92.9$}
        \\
        & & \multicolumn{1}{l}{Rec.} &
        \multicolumn{1}{l}{$91.0$} & \multicolumn{1}{l}{$91.5$} &
        \multicolumn{1}{l}{$93.0$} & \multicolumn{1}{l}{$91.0$} &
        \multicolumn{1}{l}{$94.5$} & \multicolumn{1}{l}{$95.5$} &
        \multicolumn{1}{l}{$95.0$} & \multicolumn{1}{l}{$93.0$}
        \\
        & & \multicolumn{1}{l}{Pre.} &
        \multicolumn{1}{l}{$91.0$} & \multicolumn{1}{l}{$91.8$} &
        \multicolumn{1}{l}{$93.0$} & \multicolumn{1}{l}{$91.0$} &
        \multicolumn{1}{l}{$94.5$} & \multicolumn{1}{l}{$95.5$} &
        \multicolumn{1}{l}{$95.0$} & \multicolumn{1}{l}{$93.0$}
        \\
        \cmidrule(lr){2-11}
        \multicolumn{1}{l}{ } &
        \multicolumn{1}{c}{\multirow{4}{*}{$R_z$}} &
        \multicolumn{1}{l}{Acc.} &
        \multicolumn{1}{l}{$87.0$} & \multicolumn{1}{l}{$86.0$} &
        \multicolumn{1}{l}{$88.0$} & \multicolumn{1}{l}{$\textbf{91.5}$} &
        \multicolumn{1}{l}{$91.5$} & \multicolumn{1}{l}{$88.5$} &
        \multicolumn{1}{l}{$\textbf{92.5}$} & \multicolumn{1}{l}{$91.5$}
        \\
        & & \multicolumn{1}{l}{F1} &
        \multicolumn{1}{l}{$87.0$} & \multicolumn{1}{l}{$85.9$} &
        \multicolumn{1}{l}{$87.9$} & \multicolumn{1}{l}{$91.4$} &
        \multicolumn{1}{l}{$91.4$} & \multicolumn{1}{l}{$88.4$} &
        \multicolumn{1}{l}{$92.4$} & \multicolumn{1}{l}{$91.4$}
        \\
        & & \multicolumn{1}{l}{Rec.} &
        \multicolumn{1}{l}{$87.0$} & \multicolumn{1}{l}{$86.0$} &
        \multicolumn{1}{l}{$88.0$} & \multicolumn{1}{l}{$91.5$} &
        \multicolumn{1}{l}{$91.5$} & \multicolumn{1}{l}{$88.5$} &
        \multicolumn{1}{l}{$92.5$} & \multicolumn{1}{l}{$91.5$}
        \\
        & & \multicolumn{1}{l}{Pre.} &
        \multicolumn{1}{l}{$87.0$} & \multicolumn{1}{l}{$86.2$} &
        \multicolumn{1}{l}{$88.1$} & \multicolumn{1}{l}{$91.5$} &
        \multicolumn{1}{l}{$91.8$} & \multicolumn{1}{l}{$89.1$} &
        \multicolumn{1}{l}{$92.5$} & \multicolumn{1}{l}{$91.6$}
        \\
        \bottomrule
    \end{tabular}
  \end{adjustbox}

  \label{tab:1_cifar10}
\end{table*}

\begin{table*}[ht]
    \caption{The average accuracy and standard deviation of CNN and HQCNN models across five training runs on the CIFAR-10 test set for two binary classification tasks, comparing performance across RGB, Lab, YCrCb, and HSV color spaces and HQCNN configurations.}
    \begin{adjustbox}{center}
    \centering
    \footnotesize
    \begin{tabular}{l * {11}{c}}
        \toprule
        \multicolumn{1}{l}{\multirow{1}{*}{Model}} &
        \multicolumn{1}{l}{\multirow{1}{*}{$U_{E}$}} &
        \multicolumn{1}{l}{\multirow{1}{*}{$U_{OA}$}} &
        \multicolumn{2}{l}{\multirow{1}{*}{Dataset}} &
        \multicolumn{4}{c}{\multirow{1}{*}{CIFAR-10}}
        \\
        \cmidrule(lr){4-11}
        & & & \multicolumn{1}{l}{$\mathrm{C}_{1}^\mathrm{RGB}$} & \multicolumn{1}{l}{$\mathrm{C}_{1}^\mathrm{Lab}$} &
        \multicolumn{1}{l}{$\mathrm{C}_{1}^\mathrm{YCrCb}$} & \multicolumn{1}{l}{$\mathrm{C}_{1}^\mathrm{HSV}$} &
        \multicolumn{1}{l}{$\mathrm{C}_{2}^\mathrm{RGB}$} & \multicolumn{1}{l}{$\mathrm{C}_{2}^\mathrm{Lab}$} &
        \multicolumn{1}{l}{$\mathrm{C}_{2}^\mathrm{YCrCb}$} & \multicolumn{1}{l}{$\mathrm{C}_{2}^\mathrm{HSV}$}
        \\
        \midrule
        \multicolumn{1}{l}{\multirow{1}{*}{CNN}} &
        \multicolumn{1}{l}{\multirow{5}{*}{ }} & &
        \multicolumn{1}{l}{$91.7\pm 1.5$} & \multicolumn{1}{l}{$93.0 \pm 0.9$} &
        \multicolumn{1}{l}{$\textbf{94.1} \pm \textbf{0.9}$} & \multicolumn{1}{l}{$91.1 \pm 0.9$} &
        \multicolumn{1}{l}{$93.3\pm 0.5$} & \multicolumn{1}{l}{$\textbf{94.5}\pm \textbf{0.9}$} &
        \multicolumn{1}{l}{$94.5\pm 1.0$} & \multicolumn{1}{l}{$92.5\pm 1.3$}
        \\
        \cmidrule(lr){1-11}
        \multicolumn{1}{l}{\multirow{9}{*}{HQCNN}} &
        \multicolumn{1}{c}{\multirow{3}{*}{$R_{YZ}$}} &

        \multicolumn{1}{c}{$R_{x}$} &
        \multicolumn{1}{l}{$89.6 \pm 2.0$} & \multicolumn{1}{l}{$89.9 \pm 1.9$} &
        \multicolumn{1}{l}{$\textbf{91.1} \pm \textbf{3.0}$} & \multicolumn{1}{l}{$89.7 \pm 1.4$} &
        \multicolumn{1}{l}{$92.1 \pm 1.9$} & \multicolumn{1}{l}{$91.1 \pm 1.6$} &
        \multicolumn{1}{l}{$\textbf{93.1} \pm \textbf{2.0}$} & \multicolumn{1}{l}{$90.8 \pm 2.2$}
        \\
        & & \multicolumn{1}{c}{$R_{y}$} &
        \multicolumn{1}{l}{$90.4 \pm 1.9$} & \multicolumn{1}{l}{$\textbf{91.7} \pm \textbf{1.4}$} &
        \multicolumn{1}{l}{$90.7 \pm 1.2$} & \multicolumn{1}{l}{$90.1 \pm 1.5$} &
        \multicolumn{1}{l}{$91.6 \pm 4.9$} & \multicolumn{1}{l}{$92.1 \pm 1.0$} &
        \multicolumn{1}{l}{$\textbf{92.5} \pm \textbf{1.3}$} & \multicolumn{1}{l}{$91.2 \pm 2.5$}
        \\
        & & \multicolumn{1}{c}{$R_{z}$} &
        \multicolumn{1}{l}{$86.9 \pm 4.0$} & \multicolumn{1}{l}{$81.1 \pm 3.5$} &
        \multicolumn{1}{l}{$85.4 \pm 2.4$} & \multicolumn{1}{l}{$\textbf{88.5} \pm \textbf{1.9}$} &
        \multicolumn{1}{l}{$\textbf{90.0} \pm \textbf{1.3}$} & \multicolumn{1}{l}{$82.8 \pm 0.9$} &
        \multicolumn{1}{l}{$87.4 \pm 1.4$} & \multicolumn{1}{l}{$89.7 \pm 3.0$}
        \\
        \cmidrule(lr){2-11}
        \multicolumn{1}{l}{ } &
        \multicolumn{1}{c}{\multirow{3}{*}{$R_{XZ}$}} &
        \multicolumn{1}{c}{$R_{x}$} &
        \multicolumn{1}{l}{$89.6 \pm 3.0$} & \multicolumn{1}{l}{$\textbf{93.0} \pm \textbf{0.7}$} &
        \multicolumn{1}{l}{$90.4 \pm 1.8$} & \multicolumn{1}{l}{$89.4 \pm 2.5$} &
        \multicolumn{1}{l}{$91.3 \pm 2.4$} & \multicolumn{1}{l}{$\textbf{92.9} \pm \textbf{1.0}$} &
        \multicolumn{1}{l}{$92.5 \pm 1.5$} & \multicolumn{1}{l}{$92.8 \pm 2.2$}
        \\
        & & \multicolumn{1}{c}{$R_{y}$} &
        \multicolumn{1}{l}{$88.6 \pm 1.7$} & \multicolumn{1}{l}{$89.2 \pm 1.4$} &
        \multicolumn{1}{l}{$\textbf{91.4} \pm \textbf{1.0}$} & \multicolumn{1}{l}{$89.9 \pm 1.0$} &
        \multicolumn{1}{l}{$93.1 \pm 1.1$} & \multicolumn{1}{l}{$\textbf{93.3} \pm \textbf{1.5}$} &
        \multicolumn{1}{l}{$92.7 \pm 1.6$} & \multicolumn{1}{l}{$91.2 \pm 1.3$}
        \\
        & & \multicolumn{1}{c}{$R_{z}$} &
        \multicolumn{1}{l}{$83.6 \pm 3.8$} & \multicolumn{1}{l}{$81.7 \pm 2.4$} &
        \multicolumn{1}{l}{$85.2 \pm 2.3$} & \multicolumn{1}{l}{$\textbf{88.9} \pm \textbf{1.4}$} &
        \multicolumn{1}{l}{$88.6 \pm 4.5$} & \multicolumn{1}{l}{$85.8 \pm 3.4$} &
        \multicolumn{1}{l}{$87.2 \pm 3.0$} & \multicolumn{1}{l}{$\textbf{89.9} \pm \textbf{2.2}$}
        \\
        \cmidrule(lr){2-11}
        \multicolumn{1}{l}{ } &
        \multicolumn{1}{c}{\multirow{3}{*}{$R_{XY}$}} &
        \multicolumn{1}{c}{$R_{x}$} &
        \multicolumn{1}{l}{$90.2 \pm 2.0$} & \multicolumn{1}{l}{$\textbf{91.6} \pm \textbf{1.8}$} &
        \multicolumn{1}{l}{$89.5 \pm 1.6$} & \multicolumn{1}{l}{$88.9 \pm 1.8$} &
        \multicolumn{1}{l}{$91.2 \pm 2.1$} & \multicolumn{1}{l}{$92.1 \pm 1.1$} &
        \multicolumn{1}{l}{$\textbf{92.3} \pm \textbf{1.7}$} & \multicolumn{1}{l}{$89.2 \pm 1.2$}
        \\
        & &\multicolumn{1}{c}{$R_{y}$} &
        \multicolumn{1}{l}{$90.3 \pm 1.5$} & \multicolumn{1}{l}{$\textbf{91.6} \pm \textbf{1.5}$} &
        \multicolumn{1}{l}{$91.0 \pm 1.3$} & \multicolumn{1}{l}{$90.5 \pm 1.5$} &
        \multicolumn{1}{l}{$\textbf{93.2} \pm \textbf{2.0}$} & \multicolumn{1}{l}{$92.9 \pm 1.5$} &
        \multicolumn{1}{l}{$92.4 \pm 1.2$} & \multicolumn{1}{l}{$90.8 \pm 2.6$}
        \\
        & & \multicolumn{1}{c}{$R_{z}$} &
        \multicolumn{1}{l}{$86.7 \pm 2.3$} & \multicolumn{1}{l}{$80.9 \pm 3.7$} &
        \multicolumn{1}{l}{$84.8 \pm 1.5$} & \multicolumn{1}{l}{$\textbf{88.1} \pm \textbf{2.0}$} &
        \multicolumn{1}{l}{$87.8 \pm 5.2$} & \multicolumn{1}{l}{$82.6 \pm 3.3$} &
        \multicolumn{1}{l}{$88.9 \pm 1.5$} & \multicolumn{1}{l}{$\textbf{90.6} \pm \textbf{1.3}$}
        \\
        \bottomrule
    \end{tabular}
    \end{adjustbox}

  \label{tab:2_cifar10_mean}
\end{table*}

\subsection{EuroSAT}
Performance metrics for the EuroSAT dataset, as shown in~\Cref{tab:3_eurosat}. For the $\textrm{E}_1$ (annualcrop-fores) classification task, comparable performance is observed between the CNN and HQCNN models
For the $\textrm{E}_2$ (industrial-river) task, the HQCNN using $R_x$ gate achieves an accuracy of $98.5\%$ in Lab, while HSV reaches $97.5\%$, matching the CNN's HSV performance and surpassing RGB at $96\%$. 

\begin{table*}[htb]
    \caption{
    The best accuracy ($\%$) from five training runs on the EuroSAT dataset.
    }
    \begin{adjustbox}{center}
    \centering
     \footnotesize
    \begin{tabular}{l * {11}{c}}
        \toprule
        \multicolumn{1}{l}{\multirow{1}{*}{Model}} &
        \multicolumn{2}{l}{\multirow{1}{*}{$U_{OA}$}} &
        \multicolumn{2}{l}{\multirow{1}{*}{Dataset}} &
        \multicolumn{4}{c}{\multirow{1}{*}{EuroSAT}}
        \\
        \cmidrule(lr){4-11}
        & & & \multicolumn{1}{l}{$\mathrm{E}_{1}^\mathrm{RGB\ \ }$} & \multicolumn{1}{l}{$\mathrm{E}_{1}^\mathrm{Lab\ \ }$} &
        \multicolumn{1}{l}{$\mathrm{E}_{1}^\mathrm{YCrCb}$} & \multicolumn{1}{l}{$\mathrm{E}_{1}^\mathrm{HSV\ \ }$} &
        \multicolumn{1}{l}{$\mathrm{E}_{2}^\mathrm{RGB\ \ }$} & \multicolumn{1}{l}{$\mathrm{E}_{2}^\mathrm{Lab\ \ }$} &
        \multicolumn{1}{l}{$\mathrm{E}_{2}^\mathrm{YCrCb}$} & \multicolumn{1}{l}{$\mathrm{E}_{2}^\mathrm{HSV\ \ }$}
        \\
        \midrule
        \multicolumn{1}{l}{\multirow{4}{*}{CNN}} &
        \multicolumn{1}{l}{\multirow{4}{*}{ }} &
        \multicolumn{1}{l}{Acc.} &
        \multicolumn{1}{l}{$\textbf{99.5}$} & \multicolumn{1}{l}{$\textbf{99.5}$} &
        \multicolumn{1}{l}{$99.0$} & \multicolumn{1}{l}{$\textbf{99.5}$} &
        \multicolumn{1}{l}{$98.5$} & \multicolumn{1}{l}{$\textbf{99.5}$} &
        \multicolumn{1}{l}{$\textbf{99.5}$} & \multicolumn{1}{l}{$97.5$}
        \\
        & & \multicolumn{1}{l}{F1} &
        \multicolumn{1}{l}{$99.4$} & \multicolumn{1}{l}{$99.4$} &
        \multicolumn{1}{l}{$98.9$} & \multicolumn{1}{l}{$99.4$} &
        \multicolumn{1}{l}{$98.4$} & \multicolumn{1}{l}{$99.4$} &
        \multicolumn{1}{l}{$99.4$} & \multicolumn{1}{l}{$97.4$}
        \\
        & & \multicolumn{1}{l}{Rec.} &
        \multicolumn{1}{l}{$99.5$} & \multicolumn{1}{l}{$99.5$} &
        \multicolumn{1}{l}{$99.0$} & \multicolumn{1}{l}{$99.5$} &
        \multicolumn{1}{l}{$98.5$} & \multicolumn{1}{l}{$99.5$} &
        \multicolumn{1}{l}{$99.5$} & \multicolumn{1}{l}{$97.5$}
        \\
        & & \multicolumn{1}{l}{Pre.} &
        \multicolumn{1}{l}{$99.5$} & \multicolumn{1}{l}{$99.5$} &
        \multicolumn{1}{l}{$99.0$} & \multicolumn{1}{l}{$99.5$} &
        \multicolumn{1}{l}{$98.5$} & \multicolumn{1}{l}{$99.5$} &
        \multicolumn{1}{l}{$99.5$} & \multicolumn{1}{l}{$97.5$}

        \\
        \cmidrule(lr){1-11}
        \multicolumn{1}{l}{\multirow{12}{*}{HQCNN}} &
        \multicolumn{1}{c}{\multirow{4}{*}{$R_x$}} &
        \multicolumn{1}{l}{Acc.} &
        \multicolumn{1}{l}{$99.0$} & \multicolumn{1}{l}{$99.5$} &
        \multicolumn{1}{l}{$\textbf{100}$} & \multicolumn{1}{l}{$99.5$} &
        \multicolumn{1}{l}{$96.0$} & \multicolumn{1}{l}{$\textbf{98.5}$} &
        \multicolumn{1}{l}{$96.5$} & \multicolumn{1}{l}{$97.5$}
        \\
        & & \multicolumn{1}{l}{F1} &
        \multicolumn{1}{l}{$99.0$} & \multicolumn{1}{l}{$99.4$} &
        \multicolumn{1}{l}{$100$} & \multicolumn{1}{l}{$99.4$} &
        \multicolumn{1}{l}{$96.0$} & \multicolumn{1}{l}{$98.4$} &
        \multicolumn{1}{l}{$96.4$} & \multicolumn{1}{l}{$97.4$}
        \\
        & & \multicolumn{1}{l}{Rec.} &
        \multicolumn{1}{l}{$99.0$} & \multicolumn{1}{l}{$99.5$} &
        \multicolumn{1}{l}{$100$} & \multicolumn{1}{l}{$99.5$} &
        \multicolumn{1}{l}{$96.0$} & \multicolumn{1}{l}{$98.5$} &
        \multicolumn{1}{l}{$96.5$} & \multicolumn{1}{l}{$97.5$}
        \\
        & & \multicolumn{1}{l}{Pre.} &
        \multicolumn{1}{l}{$99.0$} & \multicolumn{1}{l}{$99.5$} &
        \multicolumn{1}{l}{$100$} & \multicolumn{1}{l}{$99.5$} &
        \multicolumn{1}{l}{$96.0$} & \multicolumn{1}{l}{$98.5$} &
        \multicolumn{1}{l}{$96.5$} & \multicolumn{1}{l}{$97.5$}
        \\
        \cmidrule(lr){2-11}
        \multicolumn{1}{l}{ } &
        \multicolumn{1}{c}{\multirow{4}{*}{$R_y$}} &
        \multicolumn{1}{l}{Acc.} &
        \multicolumn{1}{l}{$\textbf{99.5}$} & \multicolumn{1}{l}{$\textbf{99.5}$} &
        \multicolumn{1}{l}{$99.0$} & \multicolumn{1}{l}{$99.0$} &
        \multicolumn{1}{l}{$97.0$} & \multicolumn{1}{l}{$\textbf{97.5}$} &
        \multicolumn{1}{l}{$\textbf{97.5}$} & \multicolumn{1}{l}{$96.5$}
        \\
        & & \multicolumn{1}{l}{F1} &
        \multicolumn{1}{l}{$99.4$} & \multicolumn{1}{l}{$99.4$} &
        \multicolumn{1}{l}{$98.9$} & \multicolumn{1}{l}{$98.9$} &
        \multicolumn{1}{l}{$96.9$} & \multicolumn{1}{l}{$97.4$} &
        \multicolumn{1}{l}{$97.4$} & \multicolumn{1}{l}{$96.4$}
        \\
        & & \multicolumn{1}{l}{Rec.} &
        \multicolumn{1}{l}{$99.5$} & \multicolumn{1}{l}{$99.5$} &
        \multicolumn{1}{l}{$99.0$} & \multicolumn{1}{l}{$99.0$} &
        \multicolumn{1}{l}{$97.0$} & \multicolumn{1}{l}{$97.5$} &
        \multicolumn{1}{l}{$97.5$} & \multicolumn{1}{l}{$96.5$}
        \\
        & & \multicolumn{1}{l}{Pre.} &
        \multicolumn{1}{l}{$99.5$} & \multicolumn{1}{l}{$99.5$} &
        \multicolumn{1}{l}{$99.0$} & \multicolumn{1}{l}{$99.0$} &
        \multicolumn{1}{l}{$97.0$} & \multicolumn{1}{l}{$97.5$} &
        \multicolumn{1}{l}{$97.5$} & \multicolumn{1}{l}{$96.5$}
        \\
        \cmidrule(lr){2-11}
        \multicolumn{1}{l}{ } &
        \multicolumn{1}{c}{\multirow{4}{*}{$R_z$}} &
        \multicolumn{1}{l}{Acc.} &
        \multicolumn{1}{l}{$\textbf{99.5}$} & \multicolumn{1}{l}{$99.0$} &
        \multicolumn{1}{l}{$99.0$} & \multicolumn{1}{l}{$99.0$} &
        \multicolumn{1}{l}{$\textbf{97.5}$} & \multicolumn{1}{l}{$96.5$} &
        \multicolumn{1}{l}{$96.5$} & \multicolumn{1}{l}{$96.5$}
        \\
        & & \multicolumn{1}{l}{F1} &
        \multicolumn{1}{l}{$99.4$} & \multicolumn{1}{l}{$98.9$} &
        \multicolumn{1}{l}{$99.0$} & \multicolumn{1}{l}{$98.9$} &
        \multicolumn{1}{l}{$97.4$} & \multicolumn{1}{l}{$96.4$} &
        \multicolumn{1}{l}{$96.4$} & \multicolumn{1}{l}{$96.4$}
        \\
        & & \multicolumn{1}{l}{Rec.} &
        \multicolumn{1}{l}{$99.5$} & \multicolumn{1}{l}{$99.0$} &
        \multicolumn{1}{l}{$99.0$} & \multicolumn{1}{l}{$99.0$} &
        \multicolumn{1}{l}{$97.5$} & \multicolumn{1}{l}{$96.5$} &
        \multicolumn{1}{l}{$96.5$} & \multicolumn{1}{l}{$96.5$}
        \\
        & & \multicolumn{1}{l}{Pre.} &
        \multicolumn{1}{l}{$99.5$} & \multicolumn{1}{l}{$99.0$} &
        \multicolumn{1}{l}{$99.0$} & \multicolumn{1}{l}{$99.0$} &
        \multicolumn{1}{l}{$97.5$} & \multicolumn{1}{l}{$96.5$} &
        \multicolumn{1}{l}{$96.5$} & \multicolumn{1}{l}{$96.5$}
        \\
        \bottomrule
    \end{tabular}
    \end{adjustbox}
  \label{tab:3_eurosat}
\end{table*}

\begin{table*}[htb]
    \caption{The average and standard deviation of results ($\%$) from five test set evaluations.}
    \begin{adjustbox}{center}
    \centering
     \footnotesize
    \begin{tabular}{l * {11}{c}}
        \toprule
        \multicolumn{1}{l}{\multirow{1}{*}{Model}} &
        \multicolumn{1}{l}{\multirow{1}{*}{$U_{E}$}} &
        \multicolumn{1}{l}{\multirow{1}{*}{$U_{OA}$}} &
        \multicolumn{2}{l}{\multirow{1}{*}{Dataset}} &
        \multicolumn{4}{c}{\multirow{1}{*}{EuroSAT}}
        \\
        \cmidrule(lr){4-11}
        & & & \multicolumn{1}{l}{$\mathrm{E}_{1}^\mathrm{RGB}$} & \multicolumn{1}{l}{$\mathrm{E}_{1}^\mathrm{Lab}$} &
        \multicolumn{1}{l}{$\mathrm{E}_{1}^\mathrm{YCrCb}$} & \multicolumn{1}{l}{$\mathrm{E}_{1}^\mathrm{HSV}$} &
        \multicolumn{1}{l}{$\mathrm{E}_{2}^\mathrm{RGB}$} & \multicolumn{1}{l}{$\mathrm{E}_{2}^\mathrm{Lab}$} &
        \multicolumn{1}{l}{$\mathrm{E}_{2}^\mathrm{YCrCb}$} & \multicolumn{1}{l}{$\mathrm{E}_{2}^\mathrm{HSV}$}
        \\
        \midrule
        \multicolumn{1}{l}{\multirow{1}{*}{CNN}} &
        \multicolumn{1}{l}{\multirow{5}{*}{ }} & &
        \multicolumn{1}{l}{$98.5 \pm 1.0$} & \multicolumn{1}{l}{$\textbf{98.9} \pm \textbf{0.5}$} &
        \multicolumn{1}{l}{$98.6 \pm 0.3$} & \multicolumn{1}{l}{$98.4 \pm 0.7$} &
        \multicolumn{1}{l}{$97.8 \pm 0.7$} & \multicolumn{1}{l}{$97.9 \pm 0.8$} &
        \multicolumn{1}{l}{$\textbf{98.6} \pm \textbf{0.4}$} & \multicolumn{1}{l}{$96.9 \pm 0.3$}

        \\
        \cmidrule(lr){1-11}
        \multicolumn{1}{l}{\multirow{9}{*}{HQCNN}} &
        \multicolumn{1}{c}{\multirow{3}{*}{$R_{YZ}$}} &

        \multicolumn{1}{c}{$R_{x}$} &
        \multicolumn{1}{l}{$\textbf{99.0} \pm \textbf{0.4}$} & \multicolumn{1}{l}{$98.6 \pm 0.7$} &
        \multicolumn{1}{l}{$98.8 \pm 0.5$} & \multicolumn{1}{l}{$98.0 \pm 0.5$} &
        \multicolumn{1}{l}{$94.7 \pm 0.8$} & \multicolumn{1}{l}{$\textbf{95.3} \pm \textbf{0.9}$} &
        \multicolumn{1}{l}{$95.1 \pm 0.8$} & \multicolumn{1}{l}{$94.8 \pm 1.9$}
        \\
        & & \multicolumn{1}{c}{$R_{y}$} &
        \multicolumn{1}{l}{$98.2 \pm 0.6$} & \multicolumn{1}{l}{$\textbf{99.1} \pm \textbf{0.5}$} &
        \multicolumn{1}{l}{$98.6 \pm 0.5$} & \multicolumn{1}{l}{$97.5 \pm 1.5$} &
        \multicolumn{1}{l}{$94.1 \pm 0.2$} & \multicolumn{1}{l}{$94.9 \pm 0.9$} &
        \multicolumn{1}{l}{$94.6 \pm 1.5$} & \multicolumn{1}{l}{$\textbf{95.6} \pm \textbf{1.1}$}
        \\
        & & \multicolumn{1}{c}{$R_{z}$} &
        \multicolumn{1}{l}{$98.5 \pm 1.1$} & \multicolumn{1}{l}{$99.0 \pm 1.0$} &
        \multicolumn{1}{l}{$98.6 \pm 1.1$} & \multicolumn{1}{l}{$\textbf{99.0} \pm \textbf{0.6}$} &
        \multicolumn{1}{l}{$\textbf{94.5} \pm \textbf{1.3}$} & \multicolumn{1}{l}{$94.0 \pm 1.4$} &
        \multicolumn{1}{l}{$94.0 \pm 1.1$} & \multicolumn{1}{l}{$94.2 \pm 1.9$}
        \\
        \cmidrule(lr){2-11}
        \multicolumn{1}{l}{ } &
        \multicolumn{1}{c}{\multirow{3}{*}{$R_{XZ}$}} &
        \multicolumn{1}{c}{$R_{x}$} &
        \multicolumn{1}{l}{$98.2 \pm 0.8$} & \multicolumn{1}{l}{$98.8 \pm 0.9$} &
        \multicolumn{1}{l}{$\textbf{99.3} \pm \textbf{0.6}$} & \multicolumn{1}{l}{$99.0 \pm 0.5$} &
        \multicolumn{1}{l}{$94.6 \pm 1.2$} & \multicolumn{1}{l}{$95.5 \pm 1.7$} &
        \multicolumn{1}{l}{$\textbf{95.8} \pm \textbf{0.5}$} & \multicolumn{1}{l}{$95.4 \pm 1.3$}
        \\
        & & \multicolumn{1}{c}{$R_{y}$} &
        \multicolumn{1}{l}{$98.6 \pm 1.1$} & \multicolumn{1}{l}{$\textbf{98.8} \pm \textbf{0.4}$} &
        \multicolumn{1}{l}{$98.7 \pm 0.2$} & \multicolumn{1}{l}{$97.5 \pm 1.5$} &
        \multicolumn{1}{l}{$95.0 \pm 1.1$} & \multicolumn{1}{l}{$94.9 \pm 1.4$} &
        \multicolumn{1}{l}{$94.8 \pm 1.8$} & \multicolumn{1}{l}{$\textbf{95.3} \pm \textbf{1.0}$}
        \\
        & & \multicolumn{1}{c}{$R_{z}$} &
        \multicolumn{1}{l}{$98.3 \pm 0.8$} & \multicolumn{1}{l}{$\textbf{98.4} \pm \textbf{0.3}$} &
        \multicolumn{1}{l}{$\textbf{98.4} \pm \textbf{0.3}$} & \multicolumn{1}{l}{$98.3 \pm 0.6$} &
        \multicolumn{1}{l}{$\textbf{96.6} \pm \textbf{0.6}$} & \multicolumn{1}{l}{$93.8 \pm 1.6$} &
        \multicolumn{1}{l}{$95.2 \pm 0.8$} & \multicolumn{1}{l}{$95.0 \pm 1.0$}
        \\
        \cmidrule(lr){2-11}
        \multicolumn{1}{l}{ } &
        \multicolumn{1}{c}{\multirow{3}{*}{$R_{XY}$}} &
        \multicolumn{1}{c}{$R_{x}$} &
        \multicolumn{1}{l}{$98.3 \pm 0.7$} & \multicolumn{1}{l}{$\textbf{98.6} \pm \textbf{0.6}$} &
        \multicolumn{1}{l}{$98.1 \pm 0.5$} & \multicolumn{1}{l}{$97.9 \pm 1.2$} &
        \multicolumn{1}{l}{$94.0 \pm 0.9$} & \multicolumn{1}{l}{$93.3 \pm 1.1$} &
        \multicolumn{1}{l}{$94.6 \pm 1.9$} & \multicolumn{1}{l}{$\textbf{95.5} \pm \textbf{0.8}$}
        \\
        & &\multicolumn{1}{c}{$R_{y}$} &
        \multicolumn{1}{l}{$98.3 \pm 0.7$} & \multicolumn{1}{l}{$98.2 \pm 0.9$} &
        \multicolumn{1}{l}{$\textbf{99.6} \pm \textbf{0.2}$} & \multicolumn{1}{l}{$98.6 \pm 0.8$} &
        \multicolumn{1}{l}{$93.7 \pm 1.6$} & \multicolumn{1}{l}{$93.4 \pm 0.9$} &
        \multicolumn{1}{l}{$\textbf{95.7} \pm \textbf{0.6}$} & \multicolumn{1}{l}{$95.7 \pm 1.6$}
        \\
        & & \multicolumn{1}{c}{$R_{z}$} &
        \multicolumn{1}{l}{$98.2 \pm 1.2$} & \multicolumn{1}{l}{$98.7 \pm 0.5$} &
        \multicolumn{1}{l}{$\textbf{98.8} \pm \textbf{0.5}$} & \multicolumn{1}{l}{$98.0 \pm 1.0$} &
        \multicolumn{1}{l}{$93.6 \pm 0.5$} & \multicolumn{1}{l}{$94.7 \pm 0.8$} &
        \multicolumn{1}{l}{$94.6 \pm 1.8$} & \multicolumn{1}{l}{$\textbf{95.5} \pm \textbf{1.1}$}
        \\
        \bottomrule
    \end{tabular}
    \end{adjustbox}
  \label{tab:4_eurosat_mean}
\end{table*}


Average accuracy are summarized in~\Cref{tab:4_eurosat_mean}. The CNN model attains an average accuracy of $98.9\%$ in Lab for $\textrm{E}_1$ and $98.6\%$ in YCrCb for $\textrm{E}_2$, both surpassing RGB’s $98.5\%$.
In the HQCNN model, the $R_{YZ}-R_x$ configuration achieves an accuracy of $99\%$ in RGB for $\textrm{E}_1$, outperforming the CNN. The $R_{XY}$-$R_y$ configuration in YCrCb achieves the highest accuracy of $99.6\%$ for $E_1$. For $\textrm{E}_2$, the HQCNN does not outperform the CNN. 
However, the $U_E$-$U_{OA}$ configurations in YCrCb and Lab generally outperform RGB for $\textrm{E}_1$, while YCrCb and HSV show higher average performance than RGB in $\textrm{E}_2$ with the $R_{XY}$ encoding.

\subsection{SAT-4}
Similar to the EuroSAT dataset, both the CNN and HQCNN models achieve high accuracy in the SAT-4 dataset for the $\textrm{S}_1$ (barrenland-grassland) and $\textrm{S}_2$ (barrenland-trees) classification tasks.
For the CNN model, an accuracy of $98\%$ is achieved in Lab for $\textrm{S}_1$. 
In the HQCNN model, the $R_y$ gate achieves an accuracy of $98.5\%$ in both RGB and YCrCb, surpassing the CNN accuracies of $95\%$ and $97.5\%$ in these color spaces. 
Both models exhibit comparable performance across various color spaces in the $\textrm{S}_2$ task, as shown in~\Cref{tab:5_sat4}.

Average performance as shown in~\Cref{tab:6_sat4_mean}. In $\textrm{S}_1$, the CNN model achieves an average accuracy of $96.7\%$ in the Lab and HSV,  while the HQCNN model achieves accuracies of $96.3\%$ ($R_{YZ}-R_x$) in RGB and $97.2\%$ ($R_{XZ}-R_y$) in YCrCb, compared to the CNN's corresponding results. 
In $\textrm{S}_2$, the CNN model achieves an average accuracy of $99.7\%$ in the Lab. The HQCNN, across various configurations, achieves an average accuracy of $99.8\%$ in each color space, outperforming the CNN. 
Unlike in the CIFAR-10 dataset, the $R_z$ gate does not exhibit significant performance differences compared to other rotation gates in these tasks.

\begin{table*}[htb]
    \caption{
     The highest accuracy ($\%$) was achieved across five training sessions on the SAT-4 dataset. 
    }
    \begin{adjustbox}{center}
    \centering
     \footnotesize
    \begin{tabular}{l * {11}{c}}
        \toprule
        \multicolumn{1}{l}{\multirow{1}{*}{Model}} &
        \multicolumn{2}{l}{\multirow{1}{*}{$U_{OA}$}} &
        \multicolumn{2}{l}{\multirow{1}{*}{Dataset}} &
        \multicolumn{4}{c}{\multirow{1}{*}{SAT-4}}
        \\
        \cmidrule(lr){4-11}
        & & & \multicolumn{1}{l}{$\mathrm{S}_{1}^\mathrm{RGB\ \ }$} & \multicolumn{1}{l}{$\mathrm{S}_{1}^\mathrm{Lab\ \ }$} &
        \multicolumn{1}{l}{$\mathrm{S}_{1}^\mathrm{YCrCb}$} & \multicolumn{1}{l}{$\mathrm{S}_{1}^\mathrm{HSV\ \ }$} &
        \multicolumn{1}{l}{$\mathrm{S}_{2}^\mathrm{RGB\ \ }$} & \multicolumn{1}{l}{$\mathrm{S}_{2}^\mathrm{Lab\ \ }$} &
        \multicolumn{1}{l}{$\mathrm{S}_{2}^\mathrm{YCrCb}$} & \multicolumn{1}{l}{$\mathrm{S}_{2}^\mathrm{HSV\ \ }$}
        \\
        \midrule
        \multicolumn{1}{l}{\multirow{4}{*}{CNN}} &
        \multicolumn{1}{l}{\multirow{4}{*}{ }} &
        \multicolumn{1}{l}{Acc.} &
        \multicolumn{1}{l}{$95.0$} & \multicolumn{1}{l}{$\textbf{98.0}$} &
        \multicolumn{1}{l}{$97.5$} & \multicolumn{1}{l}{$97.5$} &
        \multicolumn{1}{l}{$100$} & \multicolumn{1}{l}{$100$} &
        \multicolumn{1}{l}{$100$} & \multicolumn{1}{l}{$100$}
        \\
        & & \multicolumn{1}{l}{F1} &
        \multicolumn{1}{l}{$94.9$} & \multicolumn{1}{l}{$97.9$} &
        \multicolumn{1}{l}{$97.4$} & \multicolumn{1}{l}{$97.4$} &
        \multicolumn{1}{l}{$100$} & \multicolumn{1}{l}{$100$} &
        \multicolumn{1}{l}{$100$} & \multicolumn{1}{l}{$100$}
        \\
        & & \multicolumn{1}{l}{Rec.} &
        \multicolumn{1}{l}{$95.0$} & \multicolumn{1}{l}{$98.0$} &
        \multicolumn{1}{l}{$97.5$} & \multicolumn{1}{l}{$97.5$} &
        \multicolumn{1}{l}{$100$} & \multicolumn{1}{l}{$100$} &
        \multicolumn{1}{l}{$100$} & \multicolumn{1}{l}{$100$}
        \\
        & & \multicolumn{1}{l}{Pre.} &
        \multicolumn{1}{l}{$95.0$} & \multicolumn{1}{l}{$98.0$} &
        \multicolumn{1}{l}{$97.5$} & \multicolumn{1}{l}{$97.5$} &
        \multicolumn{1}{l}{$100$} & \multicolumn{1}{l}{$100$} &
        \multicolumn{1}{l}{$100$} & \multicolumn{1}{l}{$100$}

        \\
        \cmidrule(lr){1-11}
        \multicolumn{1}{l}{\multirow{12}{*}{HQCNN}} &
        \multicolumn{1}{c}{\multirow{4}{*}{$R_x$}} &
        \multicolumn{1}{l}{Acc.} &
        \multicolumn{1}{l}{$96.5$} & \multicolumn{1}{l}{$\textbf{97.5}$} &
        \multicolumn{1}{l}{$96.0$} & \multicolumn{1}{l}{$96.5$} &
        \multicolumn{1}{l}{$100$} & \multicolumn{1}{l}{$99.5$} &
        \multicolumn{1}{l}{$99.5$} & \multicolumn{1}{l}{$99.5$}
        \\
        & & \multicolumn{1}{l}{F1} &
        \multicolumn{1}{l}{$96.4$} & \multicolumn{1}{l}{$97.4$} &
        \multicolumn{1}{l}{$95.9$} & \multicolumn{1}{l}{$96.4$} &
        \multicolumn{1}{l}{$100$} & \multicolumn{1}{l}{$99.4$} &
        \multicolumn{1}{l}{$99.4$} & \multicolumn{1}{l}{$99.4$}
        \\
        & & \multicolumn{1}{l}{Rec.} &
        \multicolumn{1}{l}{$96.5$} & \multicolumn{1}{l}{$97.5$} &
        \multicolumn{1}{l}{$96.0$} & \multicolumn{1}{l}{$96.5$} &
        \multicolumn{1}{l}{$100$} & \multicolumn{1}{l}{$99.5$} &
        \multicolumn{1}{l}{$99.5$} & \multicolumn{1}{l}{$99.5$}
        \\
        & & \multicolumn{1}{l}{Pre.} &
        \multicolumn{1}{l}{$96.6$} & \multicolumn{1}{l}{$97.5$} &
        \multicolumn{1}{l}{$96.0$} & \multicolumn{1}{l}{$96.5$} &
        \multicolumn{1}{l}{$100$} & \multicolumn{1}{l}{$99.5$} &
        \multicolumn{1}{l}{$99.5$} & \multicolumn{1}{l}{$99.5$}
        \\
        \cmidrule(lr){2-11}
        \multicolumn{1}{l}{ } &
        \multicolumn{1}{c}{\multirow{4}{*}{$R_y$}} &
        \multicolumn{1}{l}{Acc.} &
        \multicolumn{1}{l}{$\textbf{98.5}$} & \multicolumn{1}{l}{$98.0$} &
        \multicolumn{1}{l}{$\textbf{98.5}$} & \multicolumn{1}{l}{$97.0$} &
        \multicolumn{1}{l}{$100$} & \multicolumn{1}{l}{$100$} &
        \multicolumn{1}{l}{$100$} & \multicolumn{1}{l}{$100$}
        \\
        & & \multicolumn{1}{l}{F1} &
        \multicolumn{1}{l}{$98.4$} & \multicolumn{1}{l}{$97.9$} &
        \multicolumn{1}{l}{$98.4$} & \multicolumn{1}{l}{$96.9$} &
        \multicolumn{1}{l}{$100$} & \multicolumn{1}{l}{$100$} &
        \multicolumn{1}{l}{$100$} & \multicolumn{1}{l}{$100$}
        \\
        & & \multicolumn{1}{l}{Rec.} &
        \multicolumn{1}{l}{$98.5$} & \multicolumn{1}{l}{$98.0$} &
        \multicolumn{1}{l}{$98.5$} & \multicolumn{1}{l}{$97.0$} &
        \multicolumn{1}{l}{$100$} & \multicolumn{1}{l}{$100$} &
        \multicolumn{1}{l}{$100$} & \multicolumn{1}{l}{$100$}
        \\
        & & \multicolumn{1}{l}{Pre.} &
        \multicolumn{1}{l}{$98.5$} & \multicolumn{1}{l}{$98.0$} &
        \multicolumn{1}{l}{$98.5$} & \multicolumn{1}{l}{$97.0$} &
        \multicolumn{1}{l}{$100$} & \multicolumn{1}{l}{$100$} &
        \multicolumn{1}{l}{$100$} & \multicolumn{1}{l}{$100$}
        \\
        \cmidrule(lr){2-11}
        \multicolumn{1}{l}{ } &
        \multicolumn{1}{c}{\multirow{4}{*}{$R_z$}} &
        \multicolumn{1}{l}{Acc.} &
        \multicolumn{1}{l}{$95.5$} & \multicolumn{1}{l}{$95.5$} &
        \multicolumn{1}{l}{$\textbf{98.0}$} & \multicolumn{1}{l}{$93.5$} &
        \multicolumn{1}{l}{$100$} & \multicolumn{1}{l}{$100$} &
        \multicolumn{1}{l}{$100$} & \multicolumn{1}{l}{$99.0$}
        \\
        & & \multicolumn{1}{l}{F1} &
        \multicolumn{1}{l}{$95.4$} & \multicolumn{1}{l}{$95.4$} &
        \multicolumn{1}{l}{$97.9$} & \multicolumn{1}{l}{$93.4$} &
        \multicolumn{1}{l}{$100$} & \multicolumn{1}{l}{$100$} &
        \multicolumn{1}{l}{$100$} & \multicolumn{1}{l}{$98.9$}
        \\
        & & \multicolumn{1}{l}{Rec.} &
        \multicolumn{1}{l}{$95.5$} & \multicolumn{1}{l}{$95.5$} &
        \multicolumn{1}{l}{$98.0$} & \multicolumn{1}{l}{$93.5$} &
        \multicolumn{1}{l}{$100$} & \multicolumn{1}{l}{$100$} &
        \multicolumn{1}{l}{$100$} & \multicolumn{1}{l}{$99.0$}
        \\
        & & \multicolumn{1}{l}{Pre.} &
        \multicolumn{1}{l}{$95.5$} & \multicolumn{1}{l}{$95.7$} &
        \multicolumn{1}{l}{$98.0$} & \multicolumn{1}{l}{$93.5$} &
        \multicolumn{1}{l}{$100$} & \multicolumn{1}{l}{$100$} &
        \multicolumn{1}{l}{$100$} & \multicolumn{1}{l}{$99.0$}
        \\
        \bottomrule
    \end{tabular}
    \end{adjustbox}
  \label{tab:5_sat4}
\end{table*}

\begin{table*}[htb]
    \caption{The average accuracy ($\%$) and standard deviation of the results of the five test sets.}
    \begin{adjustbox}{center}
    \centering
     \footnotesize
    \begin{tabular}{l * {11}{c}}
        \toprule
        \multicolumn{1}{l}{\multirow{1}{*}{Model}} &
        \multicolumn{1}{l}{\multirow{1}{*}{$U_{E}$}} &
        \multicolumn{1}{l}{\multirow{1}{*}{$U_{OA}$}} &
        \multicolumn{2}{l}{\multirow{1}{*}{Dataset}} &
        \multicolumn{4}{c}{\multirow{1}{*}{SAT-4}}
        \\
        \cmidrule(lr){4-11}
        & & & \multicolumn{1}{l}{$\mathrm{S}_{1}^\mathrm{RGB}$} & \multicolumn{1}{l}{$\mathrm{S}_{1}^\mathrm{Lab}$} &
        \multicolumn{1}{l}{$\mathrm{S}_{1}^\mathrm{YCrCb}$} & \multicolumn{1}{l}{$\mathrm{S}_{1}^\mathrm{HSV}$} &
        \multicolumn{1}{l}{$\mathrm{S}_{2}^\mathrm{RGB}$} & \multicolumn{1}{l}{$\mathrm{S}_{2}^\mathrm{Lab}$} &
        \multicolumn{1}{l}{$\mathrm{S}_{2}^\mathrm{YCrCb}$} & \multicolumn{1}{l}{$\mathrm{S}_{2}^\mathrm{HSV}$}
        \\
        \midrule
        \multicolumn{1}{l}{\multirow{1}{*}{CNN}} &
        \multicolumn{1}{l}{\multirow{5}{*}{ }} & &
        \multicolumn{1}{l}{$94.0 \pm 0.9$} & \multicolumn{1}{l}{$96.7 \pm 1.0$} &
        \multicolumn{1}{l}{$96.1 \pm 1.1$} & \multicolumn{1}{l}{$\textbf{96.7} \pm \textbf{0.6}$} &
        \multicolumn{1}{l}{$99.4 \pm 0.6$} & \multicolumn{1}{l}{$\textbf{99.7} \pm \textbf{0.2}$} &
        \multicolumn{1}{l}{$99.4 \pm 0.7$} & \multicolumn{1}{l}{$99.5 \pm 0.3$}

        \\
        \cmidrule(lr){1-11}
        \multicolumn{1}{l}{\multirow{9}{*}{HQCNN}} &
        \multicolumn{1}{c}{\multirow{3}{*}{$R_{YZ}$}} &

        \multicolumn{1}{c}{$R_{x}$} &
        \multicolumn{1}{l}{$\textbf{96.3} \pm \textbf{0.9}$} & \multicolumn{1}{l}{$95.1 \pm 1.5$} &
        \multicolumn{1}{l}{$95.9 \pm 0.6$} & \multicolumn{1}{l}{$94.4 \pm 2.2$} &
        \multicolumn{1}{l}{$99.5 \pm 0.5$} & \multicolumn{1}{l}{$\textbf{99.8} \pm \textbf{0.2}$} &
        \multicolumn{1}{l}{$99.4 \pm 0.3$} & \multicolumn{1}{l}{$\textbf{99.8} \pm \textbf{0.2}$}
        \\
        & & \multicolumn{1}{c}{$R_{y}$} &
        \multicolumn{1}{l}{$95.2 \pm 1.0$} & \multicolumn{1}{l}{$94.9 \pm 0.7$} &
        \multicolumn{1}{l}{$\textbf{96.2} \pm \textbf{1.6}$} & \multicolumn{1}{l}{$96.1 \pm 0.2$} &
        \multicolumn{1}{l}{$99.5 \pm 0.3$} & \multicolumn{1}{l}{$\textbf{99.7} \pm \textbf{0.2}$} &
        \multicolumn{1}{l}{$99.2 \pm 0.2$} & \multicolumn{1}{l}{$99.7 \pm 0.4$}
        \\
        & & \multicolumn{1}{c}{$R_{z}$} &
        \multicolumn{1}{l}{$95.1 \pm 1.1$} & \multicolumn{1}{l}{$93.6 \pm 0.7$} &
        \multicolumn{1}{l}{$\textbf{95.3} \pm \textbf{2.1}$} & \multicolumn{1}{l}{$92.3 \pm 1.1$} &
        \multicolumn{1}{l}{$\textbf{99.5} \pm \textbf{0.4}$} & \multicolumn{1}{l}{$\textbf{99.5} \pm \textbf{0.4}$} &
        \multicolumn{1}{l}{$99.1 \pm 0.2$} & \multicolumn{1}{l}{$97.9 \pm 1.2$}
        \\
        \cmidrule(lr){2-11}
        \multicolumn{1}{l}{ } &
        \multicolumn{1}{c}{\multirow{3}{*}{$R_{XZ}$}} &
        \multicolumn{1}{c}{$R_{x}$} &
        \multicolumn{1}{l}{$95.5 \pm 1.2$} & \multicolumn{1}{l}{$\textbf{96.8} \pm \textbf{0.5}$} &
        \multicolumn{1}{l}{$95.2 \pm 0.6$} & \multicolumn{1}{l}{$95.1 \pm 0.8$} &
        \multicolumn{1}{l}{$\textbf{99.4} \pm \textbf{0.4}$} & \multicolumn{1}{l}{$99.2 \pm 0.6$} &
        \multicolumn{1}{l}{$99.2 \pm 0.4$} & \multicolumn{1}{l}{$98.6 \pm 0.8$}
        \\
        & & \multicolumn{1}{c}{$R_{y}$} &
        \multicolumn{1}{l}{$95.9 \pm 2.1$} & \multicolumn{1}{l}{$95.0 \pm 2.0$} &
        \multicolumn{1}{l}{$\textbf{97.2} \pm \textbf{0.9}$} & \multicolumn{1}{l}{$94.5 \pm 1.7$} &
        \multicolumn{1}{l}{$\textbf{99.8} \pm \textbf{0.2}$} & \multicolumn{1}{l}{$99.6 \pm 0.3$} &
        \multicolumn{1}{l}{$99.4 \pm 0.3$} & \multicolumn{1}{l}{$99.3 \pm 0.9$}
        \\
        & & \multicolumn{1}{c}{$R_{z}$} &
        \multicolumn{1}{l}{$94.1 \pm 1.2$} & \multicolumn{1}{l}{$93.5 \pm 1.1$} &
        \multicolumn{1}{l}{$\textbf{96.1} \pm \textbf{1.1}$} & \multicolumn{1}{l}{$90.4 \pm 3.9$} &
        \multicolumn{1}{l}{$99.6 \pm 0.3$} & \multicolumn{1}{l}{$99.4 \pm 0.9$} &
        \multicolumn{1}{l}{$\textbf{99.8} \pm \textbf{0.2}$} & \multicolumn{1}{l}{$97.6 \pm 1.5$}
        \\
        \cmidrule(lr){2-11}
        \multicolumn{1}{l}{ } &
        \multicolumn{1}{c}{\multirow{3}{*}{$R_{XY}$}} &
        \multicolumn{1}{c}{$R_{x}$} &
        \multicolumn{1}{l}{$95.3 \pm 0.8$} & \multicolumn{1}{l}{$\textbf{95.9} \pm \textbf{1.2}$} &
        \multicolumn{1}{l}{$\textbf{95.9} \pm \textbf{1.2}$} & \multicolumn{1}{l}{$94.4 \pm 2.3$} &
        \multicolumn{1}{l}{$99.4 \pm 0.4$} & \multicolumn{1}{l}{$\textbf{99.8} \pm \textbf{0.2}$} &
        \multicolumn{1}{l}{$99.5 \pm 0.4$} & \multicolumn{1}{l}{$99.3 \pm 0.4$}
        \\
        & &\multicolumn{1}{c}{$R_{y}$} &
        \multicolumn{1}{l}{$95.2 \pm 1.9$} & \multicolumn{1}{l}{$94.6 \pm 0.4$} &
        \multicolumn{1}{l}{$\textbf{96.0} \pm \textbf{1.0}$} & \multicolumn{1}{l}{$93.9 \pm 0.8$} &
        \multicolumn{1}{l}{$99.4 \pm 0.4$} & \multicolumn{1}{l}{$\textbf{99.5} \pm \textbf{0.5}$} &
        \multicolumn{1}{l}{$99.4 \pm 0.3$} & \multicolumn{1}{l}{$99.3 \pm 0.6$}
        \\
        & & \multicolumn{1}{c}{$R_{z}$} &
        \multicolumn{1}{l}{$94.6 \pm 1.1$} & \multicolumn{1}{l}{$94.4 \pm 2.2$} &
        \multicolumn{1}{l}{$\textbf{94.9} \pm \textbf{1.0}$} & \multicolumn{1}{l}{$93.0 \pm 1.5$} &
        \multicolumn{1}{l}{$98.7 \pm 0.6$} & \multicolumn{1}{l}{$99.1 \pm 0.5$} &
        \multicolumn{1}{l}{$\textbf{99.3} \pm \textbf{0.7}$} & \multicolumn{1}{l}{$98.4 \pm 1.3$}
        \\

        \bottomrule
    \end{tabular}
    \end{adjustbox}
  \label{tab:6_sat4_mean}
\end{table*}


\subsection{MNIST}

\begin{table*}[htb]
    \caption{The performance metrics ($\%$) for the ten-class classification task on the MNIST dataset. The HQCNN applies the $R_{XZ}$ encoding configuration.}
    \begin{adjustbox}{center}
    \centering
     \footnotesize
    \begin{tabular}{l * {14}{c}}
        \toprule
        \multicolumn{2}{l}{\multirow{1}{*}{Model}} &
        \multicolumn{2}{l}{\multirow{1}{*}{$U_{OA}$}} &
        \multicolumn{2}{l}{\multirow{1}{*}{ }} &
        \multicolumn{2}{l}{\multirow{1}{*}{Dataset}} &
        \multicolumn{2}{r}{\multirow{1}{*}{MNIST}}

        \\
        \cmidrule(lr){7-14}
        & & & & & & \multicolumn{2}{l}{$\mathrm{RGB}$} & \multicolumn{2}{l}{$\mathrm{Lab}$} &
        \multicolumn{2}{l}{$\mathrm{YCrCb}$} & \multicolumn{2}{l}{$\mathrm{HSV}$}
        \\
        \midrule
        \multicolumn{2}{l}{\multirow{5}{*}{CNN}} &
        \multicolumn{2}{l}{\multirow{5}{*}{ }} &
        \multicolumn{2}{l}{Acc.} &
        \multicolumn{2}{l}{$\textbf{91.1} \pm \textbf{1.0}$} & \multicolumn{2}{l}{$90.4 \pm 0.8$} &
        \multicolumn{2}{l}{$90.3 \pm 0.7$} & \multicolumn{2}{l}{$90.8 \pm 0.5$} &
        \\
        & & & & \multicolumn{2}{l}{Best} &
        \multicolumn{2}{l}{$\textbf{92.8}$} & \multicolumn{2}{l}{$91.3$} &
        \multicolumn{2}{l}{$91.6$} & \multicolumn{2}{l}{$91.5$} &
        \\
        & & & & \multicolumn{2}{l}{F1} &
        \multicolumn{2}{l}{$92.8$} & \multicolumn{2}{l}{$91.2$} &
        \multicolumn{2}{l}{$91.5$} & \multicolumn{2}{l}{$91.4$} &
        \\
        & & & & \multicolumn{2}{l}{Rec.} &
        \multicolumn{2}{l}{$92.8$} & \multicolumn{2}{l}{$91.3$} &
        \multicolumn{2}{l}{$91.6$} & \multicolumn{2}{l}{$91.5$} &
        \\
        & & & & \multicolumn{2}{l}{Pre.} &
        \multicolumn{2}{l}{$92.9$} & \multicolumn{2}{l}{$91.3$} &
        \multicolumn{2}{l}{$91.7$} & \multicolumn{2}{l}{$91.5$} &

        \\
        \cmidrule(lr){1-14}
        \multicolumn{2}{l}{\multirow{5}{*}{HQCNN}} &
        \multicolumn{2}{c}{\multirow{5}{*}{$R_x$}} &

        \multicolumn{2}{l}{Acc.} &
        \multicolumn{2}{l}{$93.0 \pm 0.8$} & \multicolumn{2}{l}{$ \textbf{93.4} \pm \textbf{0.6}$} &
        \multicolumn{2}{l}{$93.2 \pm 1.1$} & \multicolumn{2}{l}{$93.0 \pm 0.6$} &
        \\
        & & & & \multicolumn{2}{l}{Best} &
        \multicolumn{2}{l}{$93.9$} & \multicolumn{2}{l}{$\textbf{94.3}$} &
        \multicolumn{2}{l}{$94.2$} & \multicolumn{2}{l}{$93.9$} &
        \\
        & & & & \multicolumn{2}{l}{F1} &
        \multicolumn{2}{l}{$93.9$} & \multicolumn{2}{l}{$94.3$} &
        \multicolumn{2}{l}{$94.2$} & \multicolumn{2}{l}{$93.8$} &
        \\
        & & & & \multicolumn{2}{l}{Rec.} &
        \multicolumn{2}{l}{$93.9$} & \multicolumn{2}{l}{$94.3$} &
        \multicolumn{2}{l}{$94.2$} & \multicolumn{2}{l}{$93.9$} &
        \\
        & & & & \multicolumn{2}{l}{Pre.} &
        \multicolumn{2}{l}{$93.9$} & \multicolumn{2}{l}{$94.4$} &
        \multicolumn{2}{l}{$94.2$} & \multicolumn{2}{l}{$93.9$} &
        \\
        \bottomrule
    \end{tabular}
    \end{adjustbox}
  \label{tab:7_mnist10}
\end{table*}

To ensure consistency with color datasets, a three-channel version of the MNIST dataset is employed for both training and testing, unlike prior studies that utilized single-channel grayscale images.
The results are shown in~\Cref{tab:7_mnist10}. The CNN model achieves an average accuracy of $91.1\%$ in RGB, with a best accuracy of $92.8\%$. 
The HQCNN model achieves the highest average accuracy of $93.4\%$ in Lab, with a best accuracy of $94.3\%$, and $93.9\%$ in RGB.
Furthermore, the HQCNN model consistently outperforms the CNN across all four color spaces evaluated.

\begin{figure}[htb]
  \centering
   \includegraphics[width=1\linewidth]{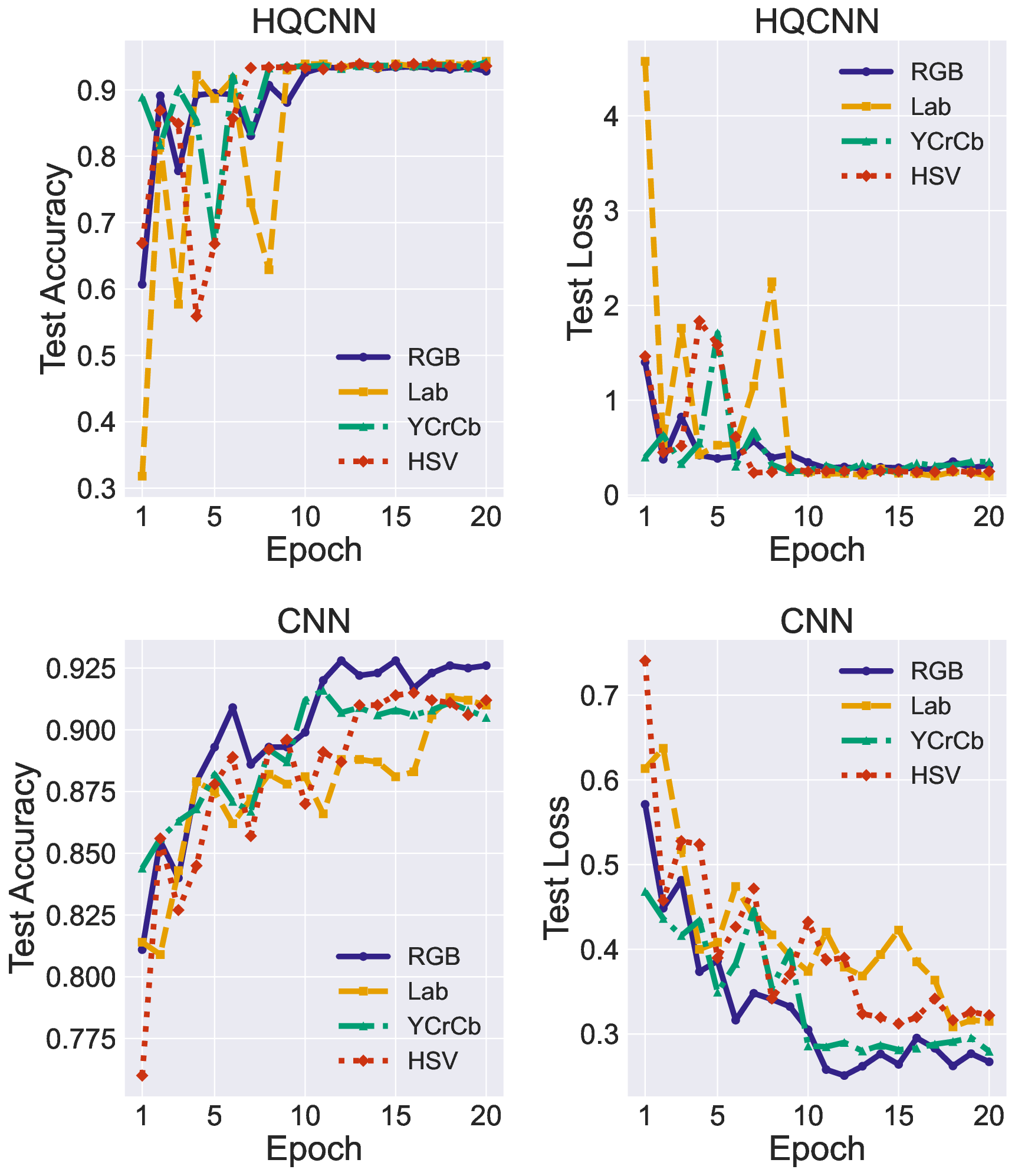}
   \caption{
        The best accuracy and loss curves of the HQCNN and the classical CNN on the MNIST test set. Data are sourced from~\Cref{tab:7_mnist10}.
   }
   \label{fig:MNIST-plot}
\end{figure}

Notably, the HQCNN demonstrates faster convergence to optimal performance, achieving stable accuracy and minimal loss within fewer epochs compared to the CNN, as shown in~\Cref{fig:MNIST-plot}.
\section{Conclusion}
\label{sec:conclusion}
This study presents an HQCNN model that integrates quantum and classical components to enhance image classification performance across multiple color spaces. 
The HQCNN demonstrates superior performance compared to a classical simple CNN baseline in several tasks, particularly in the ten-class classification task on a three-channel MNIST dataset, where it achieves the best accuracy of $94.3\%$ in the Lab color space, outperforming the CNN’s $92.8\%$ in RGB. 
Additionally, the HQCNN exhibits faster convergence, stabilizing within approximately 10 epochs of a 20-epoch training period, suggesting that quantum circuits contribute to improved efficiency in multi-channel image classification.

Across the CIFAR-10, EuroSAT, and SAT-4 datasets, the HQCNN consistently achieves competitive or superior performance in non-RGB color spaces such as Lab, YCrCb, and HSV. 
For instance, in the CIFAR-10 $\textrm{C}_2$ task, the HQCNN attains $97\%$ accuracy in HSV, surpassing the CNN’s $94.5\%$. 
In the EuroSAT $\textrm{E}_1$ task, the HQCNN reaches an average accuracy of $99.6\%$ in YCrCb with the $R_{XY}$-$R_y$ configuration, outperforming the CNN’s $98.5\%$ in RGB. 
Similarly, in the SAT-4 $\textrm{S}_1$ task, the HQCNN achieves $98.5\%$ in RGB and YCrCb with the $R_y$ gate, exceeding the CNN’s $95\%$ in RGB. 
These results suggest that non-RGB color spaces often outperform RGB, highlighting the importance of color space selection in image classification.

The experimental results demonstrate that the design of PQCs significantly influences performance.
Configurations using $R_x$ and $R_y$ gates generally outperform those with $R_z$, though $R_z$ remains competitive in specific tasks, such as those in the HSV color space.
For example, the $R_{XZ}$-$R_z$ configuration achieves competitive accuracy on CIFAR-10, demonstrating the expressiveness of optimized quantum circuits.

The hybrid quantum-classical approach enables exploration of diverse color space representations and their impact on classification performance.
However, the current study is limited by its reliance on quantum simulations rather than NISQ hardware, which may introduce practical challenges.
The varying performance of PQC designs also indicates that further optimization is needed to fully harness quantum advantages.

Future research will focus on optimizing PQC designs for hybrid systems to enhance performance. 
Evaluating the HQCNN on NISQ hardware will be essential to validate its practical applicability. 
Additionally, exploring larger datasets and other color spaces may further reveal the potential of quantum-enhanced models. 
These findings are expected to offer valuable insights for advancing quantum computing applications in computer vision, particularly in color image classification.

In summary, this study demonstrates the potential of the HQCNN to leverage quantum advantages in image classification tasks across various color spaces. 
By identifying effective color space representations and PQC configurations, the approach highlights opportunities for quantum-enhanced models in specific application scenarios, while acknowledging the need for further development to address current limitations.

\section{Acknowledgement}
This work is supported by the National Natural Science Foundation of China (Grant No. 62271234), Guangdong Basic and Applied Basic Research Foundation (Grant No. 2023A1515030290), Guangzhou Basic and Applied Basic Research Foundation (Grant No. 2025A04J3585).





%



\bibliographystyle{elsarticle-num}
\bibliography{library}

\begin{thebibliography}{10}
\expandafter\ifx\csname url\endcsname\relax
  \def\url#1{\texttt{#1}}\fi
\expandafter\ifx\csname urlprefix\endcsname\relax\def\urlprefix{URL }\fi
\expandafter\ifx\csname href\endcsname\relax
  \def\href#1#2{#2} \def\path#1{#1}\fi

\bibitem{1_de1997road}
A.~De~La~Escalera, L.~E. Moreno, M.~A. Salichs, J.~M. Armingol, Road traffic
  sign detection and classification, IEEE transactions on industrial
  electronics 44~(6) (1997) 848--859.

\bibitem{2_turay2022toward}
T.~Turay, T.~Vladimirova, Toward performing image classification and object
  detection with convolutional neural networks in autonomous driving systems: A
  survey, IEEE Access 10 (2022) 14076--14119.

\bibitem{3_song2017using}
Q.~Song, L.~Zhao, X.~Luo, X.~Dou, Using deep learning for classification of
  lung nodules on computed tomography images, Journal of healthcare engineering
  2017~(1) (2017) 8314740.

\bibitem{4_pham2020comprehensive}
T.~D. Pham, A comprehensive study on classification of covid-19 on computed
  tomography with pretrained convolutional neural networks, Scientific reports
  10~(1) (2020) 16942.

\bibitem{5_ren2022state}
Z.~Ren, F.~Fang, N.~Yan, Y.~Wu, State of the art in defect detection based on
  machine vision, International Journal of Precision Engineering and
  Manufacturing-Green Technology 9~(2) (2022) 661--691.

\bibitem{6_singh2023automated}
S.~A. Singh, K.~A. Desai, Automated surface defect detection framework using
  machine vision and convolutional neural networks, Journal of Intelligent
  Manufacturing 34~(4) (2023) 1995--2011.

\bibitem{7_yang2022development}
M.~Yang, P.~Kumar, J.~Bhola, M.~Shabaz, Development of image recognition
  software based on artificial intelligence algorithm for the efficient sorting
  of apple fruit, International Journal of System Assurance Engineering and
  Management 13~(Suppl 1) (2022) 322--330.

\bibitem{8_paymode2022transfer}
A.~S. Paymode, V.~B. Malode, Transfer learning for multi-crop leaf disease
  image classification using convolutional neural network vgg, Artificial
  Intelligence in Agriculture 6 (2022) 23--33.

\bibitem{9_rawat2017deep}
W.~Rawat, Z.~Wang, Deep convolutional neural networks for image classification:
  A comprehensive review, Neural computation 29~(9) (2017) 2352--2449.

\bibitem{10_voulodimos2018deep}
A.~Voulodimos, N.~Doulamis, A.~Doulamis, E.~Protopapadakis, Deep learning for
  computer vision: A brief review, Computational intelligence and neuroscience
  2018~(1) (2018) 7068349.

\bibitem{11_he2016deep}
K.~He, X.~Zhang, S.~Ren, J.~Sun, Deep residual learning for image recognition,
  in: Proceedings of the IEEE conference on computer vision and pattern
  recognition, 2016, pp. 770--778.

\bibitem{12_huang2017densely}
G.~Huang, Z.~Liu, L.~Van Der~Maaten, K.~Q. Weinberger, Densely connected
  convolutional networks, in: Proceedings of the IEEE conference on computer
  vision and pattern recognition, 2017, pp. 4700--4708.

\bibitem{13_jozsa2003role}
R.~Jozsa, N.~Linden, On the role of entanglement in quantum-computational
  speed-up, Proceedings of the Royal Society of London. Series A: Mathematical,
  Physical and Engineering Sciences 459~(2036) (2003) 2011--2032.

\bibitem{14_childs2003exponential}
A.~M. Childs, R.~Cleve, E.~Deotto, E.~Farhi, S.~Gutmann, D.~A. Spielman,
  Exponential algorithmic speedup by a quantum walk, in: Proceedings of the
  thirty-fifth annual ACM symposium on Theory of computing, 2003, pp. 59--68.

\bibitem{15_babbush2021focus}
R.~Babbush, J.~R. McClean, M.~Newman, C.~Gidney, S.~Boixo, H.~Neven, Focus
  beyond quadratic speedups for error-corrected quantum advantage, PRX quantum
  2~(1) (2021) 010103.

\bibitem{16_schuld2014quest}
M.~Schuld, I.~Sinayskiy, F.~Petruccione, The quest for a quantum neural
  network, Quantum Information Processing 13 (2014) 2567--2586.

\bibitem{17_gil2024understanding}
E.~Gil-Fuster, J.~Eisert, C.~Bravo-Prieto, Understanding quantum machine
  learning also requires rethinking generalization, Nature Communications
  15~(1) (2024) 2277.

\bibitem{17_1gonon2025universal}
L.~Gonon, A.~Jacquier, Universal approximation theorem and error bounds for
  quantum neural networks and quantum reservoirs, IEEE Transactions on Neural
  Networks and Learning Systems (2025) 1--14\href
  {https://doi.org/10.1109/TNNLS.2025.3552223}
  {\path{doi:10.1109/TNNLS.2025.3552223}}.

\bibitem{18_cong2019quantum}
I.~Cong, S.~Choi, M.~D. Lukin, Quantum convolutional neural networks, Nature
  Physics 15~(12) (2019) 1273--1278.

\bibitem{20_pesah2021absence}
A.~Pesah, M.~Cerezo, S.~Wang, T.~Volkoff, A.~T. Sornborger, P.~J. Coles,
  Absence of barren plateaus in quantum convolutional neural networks, Physical
  Review X 11~(4) (2021) 041011.

\bibitem{31_senokosov2024quantum}
A.~Senokosov, A.~Sedykh, A.~Sagingalieva, B.~Kyriacou, A.~Melnikov, Quantum
  machine learning for image classification, Machine Learning: Science and
  Technology 5~(1) (2024) 015040.

\bibitem{21_khanramaki2021citrus}
M.~Khanramaki, E.~A. Asli-Ardeh, E.~Kozegar, Citrus pests classification using
  an ensemble of deep learning models, Computers and Electronics in Agriculture
  186 (2021) 106192.

\bibitem{22_phan2019preserving}
H.~Phan-Xuan, T.~Le-Tien, T.~Nguyen-Chinh, T.~Do-Tieu, Q.~Nguyen-Van,
  T.~Nguyen-Thanh, Preserving spatial information to enhance performance of
  image forgery classification, in: 2019 International Conference on Advanced
  Technologies for Communications (ATC), IEEE, 2019, pp. 50--55.

\bibitem{23_moreira2022benchmark}
G.~Moreira, S.~A. Magalh{\~a}es, T.~Pinho, F.~N. dos Santos, M.~Cunha,
  Benchmark of deep learning and a proposed hsv colour space models for the
  detection and classification of greenhouse tomato, Agronomy 12~(2) (2022)
  356.

\bibitem{23_1gowda2018colornet}
S.~N. Gowda, C.~Yuan, Colornet: Investigating the importance of color spaces
  for image classification, in: Asian conference on computer vision, Springer,
  2018, pp. 581--596.

\bibitem{24_benedetti2019parameterized}
M.~Benedetti, E.~Lloyd, S.~Sack, M.~Fiorentini, Parameterized quantum circuits
  as machine learning models, Quantum Science and Technology 4~(4) (2019)
  043001.

\bibitem{25_peruzzo2014variational}
A.~Peruzzo, J.~McClean, P.~Shadbolt, M.-H. Yung, X.-Q. Zhou, P.~J. Love,
  A.~Aspuru-Guzik, J.~L. O’brien, A variational eigenvalue solver on a
  photonic quantum processor, Nature communications 5~(1) (2014) 4213.

\bibitem{26_farhi2014quantum}
E.~Farhi, J.~Goldstone, S.~Gutmann, A quantum approximate optimization
  algorithm, arXiv preprint arXiv:1411.4028 (2014).

\bibitem{27_kandala2017hardware}
A.~Kandala, A.~Mezzacapo, K.~Temme, M.~Takita, M.~Brink, J.~M. Chow, J.~M.
  Gambetta, Hardware-efficient variational quantum eigensolver for small
  molecules and quantum magnets, nature 549~(7671) (2017) 242--246.

\bibitem{28_wang2018quantum}
Z.~Wang, S.~Hadfield, Z.~Jiang, E.~G. Rieffel, Quantum approximate optimization
  algorithm for maxcut: A fermionic view, Physical Review A 97~(2) (2018)
  022304.

\bibitem{40_henderson2020quanvolutional}
M.~Henderson, S.~Shakya, S.~Pradhan, T.~Cook, Quanvolutional neural networks:
  powering image recognition with quantum circuits, Quantum Machine
  Intelligence 2~(1) (2020) 2.

\bibitem{29_ovalle2023quantum}
E.~Ovalle-Magallanes, D.~E. Alvarado-Carrillo, J.~G. Avina-Cervantes,
  I.~Cruz-Aceves, J.~Ruiz-Pinales, Quantum angle encoding with learnable
  rotation applied to quantum--classical convolutional neural networks, Applied
  Soft Computing 141 (2023) 110307.

\bibitem{30_liu2021hybrid}
J.~Liu, K.~H. Lim, K.~L. Wood, W.~Huang, C.~Guo, H.-L. Huang, Hybrid
  quantum-classical convolutional neural networks, Science China Physics,
  Mechanics \& Astronomy 64~(9) (2021) 290311.

\bibitem{32_jing2022rgb}
Y.~Jing, X.~Li, Y.~Yang, C.~Wu, W.~Fu, W.~Hu, Y.~Li, H.~Xu, Rgb image
  classification with quantum convolutional ansatz, Quantum Information
  Processing 21~(3) (2022) 101.

\bibitem{33_riaz2023accurate}
F.~Riaz, S.~Abdulla, H.~Suzuki, S.~Ganguly, R.~C. Deo, S.~Hopkins, Accurate
  image multi-class classification neural network model with quantum
  entanglement approach, Sensors 23~(5) (2023) 2753.

\bibitem{34_smaldone2023quantum}
A.~M. Smaldone, G.~W. Kyro, V.~S. Batista, Quantum convolutional neural
  networks for multi-channel supervised learning, Quantum Machine Intelligence
  5~(2) (2023) 41.

\bibitem{41_yang2022semiconductor}
Y.-F. Yang, M.~Sun, Semiconductor defect detection by hybrid classical-quantum
  deep learning, in: Proceedings of the IEEE/CVF Conference on Computer Vision
  and Pattern Recognition, 2022, pp. 2323--2332.

\bibitem{43_sim2019expressibility}
S.~Sim, P.~D. Johnson, A.~Aspuru-Guzik, Expressibility and entangling
  capability of parameterized quantum circuits for hybrid quantum-classical
  algorithms, Advanced Quantum Technologies 2~(12) (2019) 1900070.

\bibitem{39_kharsa2023advances}
R.~Kharsa, A.~Bouridane, A.~Amira, Advances in quantum machine learning and
  deep learning for image classification: a survey, Neurocomputing 560 (2023)
  126843.

\bibitem{19_cerezo2021cost}
M.~Cerezo, A.~Sone, T.~Volkoff, L.~Cincio, P.~J. Coles, Cost function dependent
  barren plateaus in shallow parametrized quantum circuits, Nature
  communications 12~(1) (2021) 1791.

\bibitem{42_barenco1995elementary}
A.~Barenco, C.~H. Bennett, R.~Cleve, D.~P. DiVincenzo, N.~Margolus, P.~Shor,
  T.~Sleator, J.~A. Smolin, H.~Weinfurter, Elementary gates for quantum
  computation, Physical review A 52~(5) (1995) 3457.

\bibitem{35_maas2013rectifier}
A.~L. Maas, A.~Y. Hannun, A.~Y. Ng, et~al., Rectifier nonlinearities improve
  neural network acoustic models, in: Proc. icml, Vol.~30, Atlanta, GA, 2013,
  p.~3.

\bibitem{36_williams1989learning}
R.~J. Williams, D.~Zipser, A learning algorithm for continually running fully
  recurrent neural networks, Neural computation 1~(2) (1989) 270--280.

\bibitem{37_rumelhart1986learning}
D.~E. Rumelhart, G.~E. Hinton, R.~J. Williams, Learning internal
  representations by error propagation, parallel distributed processing,
  explorations in the microstructure of cognition, ed. de rumelhart and j.
  mcclelland. vol. 1. 1986, Biometrika 71~(599-607) (1986) 6.

\bibitem{44_krizhevsky2009learning}
A.~Krizhevsky, G.~Hinton, et~al., Learning multiple layers of features from
  tiny images [dataset] (2009).

\bibitem{45_helber2019eurosat}
P.~Helber, B.~Bischke, A.~Dengel, D.~Borth, Eurosat: A novel dataset and deep
  learning benchmark for land use and land cover classification [dataset], IEEE
  Journal of Selected Topics in Applied Earth Observations and Remote Sensing
  12~(7) (2019) 2217--2226.

\bibitem{46_basu2015deepsat}
S.~Basu, S.~Ganguly, S.~Mukhopadhyay, R.~DiBiano, M.~Karki, R.~Nemani, Deepsat:
  a learning framework for satellite imagery [dataset], in: Proceedings of the
  23rd SIGSPATIAL international conference on advances in geographic
  information systems, 2015, pp. 1--10.

\bibitem{47_deng2012mnist}
L.~Deng, The mnist database of handwritten digit images for machine learning
  research [best of the web] [dataset], IEEE signal processing magazine 29~(6)
  (2012) 141--142.

\bibitem{48_bradski2000opencv}
G.~Bradski, A.~Kaehler, et~al., Opencv, Dr. Dobb’s journal of software tools
  3~(2) (2000).

\bibitem{49_loshchilov2017decoupled}
I.~Loshchilov, Decoupled weight decay regularization, arXiv preprint
  arXiv:1711.05101 (2017).

\bibitem{50_bergholm2018pennylane}
V.~Bergholm, J.~Izaac, M.~Schuld, C.~Gogolin, S.~Ahmed, V.~Ajith, M.~S. Alam,
  G.~Alonso-Linaje, B.~AkashNarayanan, A.~Asadi, et~al., Pennylane: Automatic
  differentiation of hybrid quantum-classical computations, arXiv preprint
  arXiv:1811.04968 (2018).

\bibitem{51_paszke2019pytorch}
A.~Paszke, S.~Gross, F.~Massa, A.~Lerer, J.~Bradbury, G.~Chanan, T.~Killeen,
  Z.~Lin, N.~Gimelshein, L.~Antiga, A.~Desmaison, A.~Kopf, E.~Yang, Z.~DeVito,
  M.~Raison, A.~Tejani, S.~Chilamkurthy, B.~Steiner, L.~Fang, J.~Bai,
  S.~Chintala, Pytorch: An imperative style, high-performance deep learning
  library, in: H.~Wallach, H.~Larochelle, A.~Beygelzimer, F.~d\textquotesingle
  Alch\'{e}-Buc, E.~Fox, R.~Garnett (Eds.), Advances in Neural Information
  Processing Systems, Vol.~32, Curran Associates, Inc., 2019.

\bibitem{52_wierichs2022general}
D.~Wierichs, J.~Izaac, C.~Wang, C.~Y.-Y. Lin, General parameter-shift rules for
  quantum gradients, Quantum 6 (2022) 677.

\end{thebibliography}






\end{document}